# Microstructure-dependent magnetic properties of iron silicon single, bi- and oligo-crystals measured with a miniaturised Single-Sheet-Tester

M. Heller[a,*], N. Leuning[b], M. Reher[a], K. Hameyer[b], S. Korte-Kerzel[a]

[a]Institute of Physical Metallurgy and Materials Physics, RWTH Aachen University, Aachen, Germany

[b]Institute of Electrical Machines, RWTH Aachen University, Aachen, Germany

*Corresponding author – Email address: heller@imm.rwth-aachen.de

## Abstract

Electrical iron silicon steel is the most commonly used soft magnetic material in electrical energy conversion and transmission, and its demand is expected to increase with the need for electrification of the transportation sector and the transition to renewable energy to combat climate change. Although iron silicon steel has been used for more than 100 years, some fundamental relationships between microstructure and magnetic performance remain vague, especially with regard to the role of crystal defects such as grain boundaries and dislocations that are induced during the final cutting step of the process chain. In this paper we present first results of a new approach to quantify the effects of orientation, grain boundaries and deformation on the magnetic properties of single, bi- and oligo-crystals using a miniaturised Single-Sheet-Tester. In this way, we were able to better resolve the orientation-dependent polarisation curves at low field strengths, revealing an additional intersection between the *medium* and *hard axis*.

Furthermore, we were able to distinguish the effects of different deformation structures – from single dislocations to tangles to localised deformation and twins – on different magnetic properties such as on coercivity, remanence and susceptibility, and we found that our particular grain boundary strongly reduces the remanence.

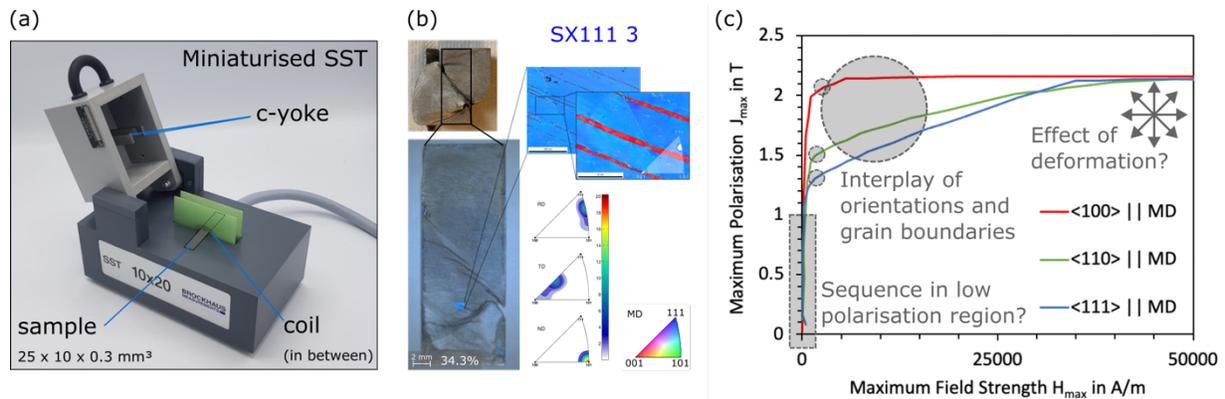

*Graphical Abstract 1: (a) Miniaturised Single-Sheet-Tester (SST) for a minimum sample size of 25 x 10 x 0.3 mm³ [1]; (b) exemplary grown and subsequently deformed (3) single crystal (SX) with an <111> axis parallel to the magnetisation direction (MD) showing localised deformation, cracks and twins; (c) state of the art polarisation curves of single crystals [2] with indicated open questions.*

# 1. Introduction

Electrical steel is one of the most important ferromagnetic materials in today's society. It is crucial both for the transmission of electric power and for the conversion of electrical energy into mechanical energy (electric motor) or vice versa (generator). That is why electrical steel is found everywhere, from the power grid and power plants to electric vehicles and kitchen appliances. Furthermore, in view of climate change and the necessity to transition to renewable energies and higher efficiencies in the use of electric energy, the demand for tailor-made, application-specific electrical steel has increased [3, 4]. To meet this demand, a deep fundamental understanding of the relationship between the microstructural features of a material and its magnetic properties is crucial.

In the production of thin electrical steel sheets with intricate cut-outs for their final application, the key material parameters interacting with the magnetic processes are changed in every process step: during rolling, grain size and dislocation density are altered through orders of magnitude with an ultimately small grain size and high dislocation density after cold rolling; during subsequent annealing, the final recrystallization texture is established and the grain size is adjusted to typically just below the order of magnitude of the sheet thickness, while the dislocation density reaches a minimum; only to be drastically increased again around the edges of all cut outs that are introduced in the final shaping step [5]. In areas where the grain diameter approaches the sheet thickness, this locally high dislocation density may occur under conditions close to single or bi-crystal deformation, with dislocations interacting strongly at and across the present grain boundaries. In the present work, we want to provide insights into the fundamental relationships between crystal orientations, deformation structures and grain boundaries on the one hand and magnetic properties on the other hand, by means of macroscopic single, bi- and oligo-crystal experiments. In particular, the literature lacks a detailed description of the influence of individual grain boundaries on magnetic properties. The foundation for today's physical understanding of ferromagnetism was laid in the early twentieth century, culminating in the Stoner model, which is still used as a starting point today. Early theoretical contributions have been made by famous scientists like Weiss et. al [6, 7], Heisenberg et. al [8], and Stoner et. al [9]. Any form of magnetism can be traced back to the quantum mechanical description of electron spins, which, in simplified terms, can be seen as tiny magnets. Atoms at the next level up can become magnetic if there is an imbalance in the electron distribution, for example, if one of the outer shells is not completely filled. However, a collection of strongly magnetic atoms can still result in an antiferromagnetic material, such as chrome, when the magnetic moments of neighbouring atoms cancel each other out. Finally, we reach the level of collections of collections of atoms (domains) in which the atomic moments are similar, but also here neighbouring collections can still cancel each other out resulting in a net zero magnetic field. To explain the amplifying effect of ferromagnetism, the Pauli exclusion principle [10] and the exchange energy are needed [8]. Let us take iron as an example. The outer electrons of individual iron atoms can freely move in the valence band, however, no two electrons located at the same atom can have the exact same quantum state (Pauli exclusion principle). One way to prevent this is to lift one electron to an energetically higher band (amplification). This, on the other hand, is only possible if the energy difference of a spin-up/spin-down configuration and a spin-up/spin-up configuration, also called exchange energy, is higher than the kinetic energy difference of the lifted and non-lifted electron state, which is only fulfilled for very few materials and is related to the atomic distance (Coulomb interaction).

Each ferromagnetic material has easier and harder to magnetize crystal directions, which is related to the optimal atomic distance for the interactions described above (spin-orbit coupling). Most experiments dealing with the fundamental physical relationships between orientation and magnetic properties were carried out on single crystals in the first half of the 20[th] century. With regard to bcc iron, the work of Beck [11], Honda and Kaya [2] as well as Webster [12] should be mentioned. They found that in bcc iron single crystals the magnetisation along <100> axes is "easier" than along "medium" <101> or "hard" <111> axes. Usually, magnetic moments align along *easy* <100> *axes*. While magnetisation along an *easy axis* goes straight to saturation, magnetisation along <101> or <111> shows a "knee" at ~1.5 T and ~1.3 T, respectively (Graphical Abstract 1 (c)). Above the knee, we are in the reversible axes deviation region. The knee polarisation values correspond to the product of maximum saturation times $\frac{1}{\sqrt{2}}$ for <101> and times $\frac{1}{\sqrt{3}}$ for <111>, which can be associated to the geometrical relationship between the axis parallel to the magnetisation direction (MD) and the next *easy axis* [13]. After these knees, the permeability decreases for both axis families, however, <111> has a slightly higher permeability than <101>, resulting in a crossing at ~2 T before both curves reach saturation (Graphical Abstract 1 (c)).

Hereafter, we will discuss the relationships between orientation, grain boundaries and deformation on the one hand and magnetic properties on the other using domain theory, which starts at the level of collections of collections of atoms [14]. Individual domains are characterised by a long-range magnetic order, with $10^{12}$ to $10^{15}$ atoms having the same magnetic moment [15]. Ferromagnetic materials exhibit magnetic hysteresis during a magnetisation cycle. At low to medium polarisations (for polycrystalline bcc iron up to approx. 1 T), domains grow through the movement (bowing and translation) of domain walls (domain growth region) [16]. In this process, slightly misaligned magnetic moments at the edge of a domain wall adjacent to a favourably aligned domain start to rotate, resulting in a displacement of the domain wall. The keyword here is exchange-anisotropy equilibrium (described below), which is slightly shifted by the external field. These movements can be reversible if there are few, weak pinning sites and irreversible if there are many, strong pinning sites. At medium to high fields (for polycrystalline bcc iron ~0.5 to ~1.5 T), entire domains with poorly aligned magnetic moments relative to the external field irreversibly rotate to a better aligned *easy axis*, overcoming the anisotropy energy (domain rotation region) [17]. Finally, at very high fields (for polycrystalline bcc iron ≥1.5 T), moments reversibly deviate from *easy axes* to align more closely with the external field (axes deviation region) [15]. The irreversible domain processes are mainly responsible for the hysteresis. While at low to medium polarisations it is mainly the microstructure that determines the course of hysteresis, at high polarisations it is the texture that dominates [18].

Without an external field, new domains form until the additional energy provided by a new domain wall is greater than the energy saved by reducing the magnetostatic energy [19]. Due to the *easy axes* distribution in bcc iron, 180° and 90° domain walls are typical. The thickness of domain walls, in which the magnetic moments change continuously, depends on the ratio between exchange and anisotropy energy of the material [15, 20]. For ferromagnetic materials, non-parallel or non-anti-parallel magnetic moments increase the exchange energy and magnetic moments that are not aligned along *easy axes* increase the anisotropy energy. Therefore, high exchange energies favour thick walls because the misalignment between the individual magnetic moments is then small, and high anisotropy energies favour thin walls because there are then few "badly" aligned moments. Through the reversible or irreversible movement of these walls, domains can grow or rotate. Impurities, microstrains and grain boundaries, among other features, can pin domain walls [15, 16, 21, 22]. When the pinning force is strong and the domain wall surface energy is low, pronounced bowing can also be observed [23, 24].

Deformation strongly affects the magnetic properties, both through magnetoelastic coupling and through the generation of dislocation structures that serve as pinning sites for domain wall movements. Regarding the former, upon magnetisation iron elastically expands in magnetisation direction as it has a positive <100> magnetostriction coefficient, which results in a tetragonal distortion of the unit cell. In [13] Webster summarises differences in magnetostriction for the different directions in iron. An increasing magnetisation along <100> results in an increasing positive magnetostriction. For <111>, all *easy axes* have the same distance resulting in a net zero magnetostriction up to the knee. However, above the knee, a strongly increasing negative magnetostriction is present when the magnetic moments are forced to deviate from the *easy axes*. <101> is a mix of both. Below the knee, part of the positive <100> magnetostriction is visible because there is always a <100> axis to rotate to that is closest to the respective <101> axis and the magnetisation direction. Above the knee, when the deviation from the *easy axes* begins, the slope decreases before becoming negative and finally ending in a negative magnetostriction at high polarisations. Once more, all these orientation dependent magnetic properties are heavily dependent on the crystal structure and atomic distance, for example, the *easy axes* from nickel are the <111> axes and all magnetostriction coefficients are negative [13, 25]. Since magnetostriction works in both directions (elongation vs. magnetisation), it is easy to imagine that local as well as global (residual) compressive stresses along the magnetisation direction heavily deteriorate the magnetic properties of iron, while tensile (residual) stresses along the magnetisation direction, at least for low values, improve the magnetic properties [26-28]. Dislocations, on the other hand, combine this magnetoelastic effect on a local microscopic level with the fact that they can act as pinning sites for domain wall movements [29]. Depending on the dislocation configuration – from individual dislocation over tangles to networks and grain boundaries – the magnitude of the pinning effect will vary. For example, B. Astié et al. [30] found that isolated, homogenously distributed dislocations hardly affect the hysteresis loss, while dislocation tangles heavily increase the hysteresis loss and after the formation of dislocation walls, the slope of deterioration decreases again. This was explained by the idea that at low dislocation densities the effect of isolated dislocations, which nevertheless act as pinning sites (especially for 90° domain walls), is dominated by the effect that pre-existing grain boundaries have on the movement of 180° domain walls, whose movements largely determine the magnetisation behaviour at low polarisations [31]. However, whether the grain boundary character is also a relevant factor remains mostly unknown. As the dislocation density increases and tangles are formed, the pinning strength increases and becomes stronger than that of grain boundaries, heavily increasing the hysteresis loss. At even higher dislocation densities, where dislocation walls start to form, the mean free path without dislocations becomes larger again, resulting in a reduced rate of hysteresis loss increase.

Looking at the above introduction, it becomes clear that the interplay between magnetic properties and different microstructural elements is very complex and needs to be understood on a fundamental basis to improve and enable better microstructure design and process engineering. Therefore, we will explore the correlation between isolated microstructural elements (crystal orientation, deformation (dislocation) structures and grain boundaries) and magnetic properties in the next sections using a new experimental approach based on targeted macroscopic crystal growth, cylinder compression, in-depth microstructure analysis and magnetic property testing with a specially designed, miniaturised Single-Sheet-Tester (SST).

## 2. Experimental

### 2.1. Material

Table 1 shows all investigated samples as well as their quantity (#), origin, and chemical composition. Industrial fully finished (FF), grain-oriented (GO) steel sheets cut in different directions (0° – parallel to RD, 45°, 90°) relative to the rolling direction (RD) were used as reference material. The single crystalline (SX) and bi-crystalline (BX) samples were grown in a custom-built, induction-based, vertical Bridgeman-Stockbarger furnace. During crystal growth the maximum current output was 260 A, the growth speed was 0.12 mm/h, and the stage was rotated with 6 rpm to guarantee a homogeneous induction field. For crystal growth, extruded polycrystalline wire material, here also investigated in its raw form (Poly), was used. All samples except the GO samples initially had a cylindric geometry, however, to test the magnetic properties cuboids with the minimum dimensions of 25 x 10 x 0.3 mm$^3$ are required. Therefore, cuboids close to these dimensions were cut via wire EDM from these cylinders, whereby the length may deviate upwards, and the final thickness was reached through etching with nitric acid (100ml HNO$_3$, 150ml H$_2$O), grinding and mechanical polishing to get rid of possible contaminations. The final surface finish for X-ray diffraction (XRD), hardness measurements and optical microscopy was achieved by a final mechanical polishing with a 1 µm diamond suspension, while an additional electropolish (A2 without water for 15 s at 24 V) was used before electron backscatter diffraction (EBSD).

*Table 1: Overview of all samples examined showing the sample name, number of analysed samples (#), origin (FF – industrial fully finished material, CG – crystal growth, raw – extruded crystal growth starting material) and chemical composition. Note: SX – single crystal, BX – bi-crystal, GO – grain oriented, Poly – polycrystal.*

| Sample | # | Origin | Si | C | Mn | P | S | Al | N | Sn | Fe |
|---|---|---|---|---|---|---|---|---|---|---|---|
| GO 0° | 6 | FF | 3.15 | 0.054 | 0.08 | 0.0215 | 0.008 | 0.02 | 0.008 | 0.08 | balance |
| GO 45° | 6 | FF | 3.15 | 0.054 | 0.08 | 0.0215 | 0.008 | 0.02 | 0.008 | 0.08 | balance |
| GO 90° | 6 | FF | 3.15 | 0.054 | 0.08 | 0.0215 | 0.008 | 0.02 | 0.008 | 0.08 | balance |
| SX100 | 4 | CG | 2.38 | 0.002 | 0.17 | 0.0012 | 0.001 | 0.02 | 0.002 | - | balance |
| SX101 | 4 | CG | 2.38 | 0.002 | 0.17 | 0.0012 | 0.001 | 0.02 | 0.002 | - | balance |
| SX111 | 4 | CG | 2.38 | 0.002 | 0.17 | 0.0012 | 0.001 | 0.02 | 0.002 | - | balance |
| BX | 4 | CG | 2.38 | 0.002 | 0.17 | 0.0012 | 0.001 | 0.02 | 0.002 | - | balance |
| Poly | 4 | raw | 2.38 | 0.002 | 0.17 | 0.0012 | 0.001 | 0.02 | 0.002 | - | balance |

Table 2 gives an overview of the different deformation states as well as orientations, present grain boundaries and angles between the respective samples' closest *easy axis* relative to MD. A more detailed description follows in section 3.

*Table 2: Overview of the samples' different deformation states (0-3), mean orientations (Bunge Euler angles), present grain boundaries (axis-angle) and minimum angle between magnetisation direction (MD) and closest easy axis. The letter after BX indicates which grain is considered in the respective line.*

| Sample | Deformation | | | | Orientation | | | Grain Boundaries | | Min. angle to MD |
|---|---|---|---|---|---|---|---|---|---|---|
| | 0 | 1 | 2 | 3 | φ1 | Φ | φ2 | Angle | Axis | |
| GO 0° | 0% | - | - | - | 264° | 36° | 94° | 17° 15° | (15 -4 11) (-16 23 -4) | 3.6° - 6.7° (13.2°) |
| GO 45° | 0% | - | - | - | 170° | 53° | 179° | 24° 11° 9° | (1 1 -2) (-1 20 -12) (21-3) | 38.3° - 43.5° |
| GO 90° | 0% | - | - | - | 221° | 45° | 180° | 10° 15° 17° | (19 17 7) (-18 7 -8) (21 14 17) | 41° - 51.9° |
| SX100 | 0% | 11.2% | 23.5% | 35.3% | 346° | 10° | 19° | 14° 8° | (-17 -11 -18) (2 27 -10) | 4° - 5.7° |
| SX101 | 0% | 11.4% | 22.2% | 33% | 195° | 25° | 206° | - | - | 40.1° |
| SX111 | 0% | 11.3% | 22.9% | 34.3% | 229° | 48° | 101° | - | - | 51.6° |
| BXa | 0% | 11.6% | 22.7% | 33.8% | 193° | 3° | 167° | (a/b) 41° | (4 9 -12) | 0.7° |
| BXb | 0% | 11.6% | 22.7% | 33.8% | 55° | 23° | 272° | (b/a') 56° | (-9 -7 4) | 39.3° |
| BXa' | 0% | 11.6% | 22.7% | 33.8% | 183° | 44° | 175° | (a/a') 41° | (-12 -1 -1) | 3.5° |
| Poly | 0% | 12.8% | 25.9% | 37.5% | - | - | - | - | - | - |

## 2.2. Compression

All in all, we investigated four grown cylinders – three single crystals and one bi-crystal – as well as one as extruded polycrystalline cylinder, all of which had a diameter of 15 mm and a length of above 50 mm. To investigate the influence of the deformation on the magnetic properties, these cylinders were deformed in several steps. First, the initial, non-deformed state (0%) was cut out of the respective cylinders. Therefore, looking at the cylinder base, a circle section remained that was then compressed, thus, the compression direction was along MD. This procedure of slicing and compressing was repeated until the four different deformation states per cylinder depicted in Table 2 were reached. All room temperature compression tests were carried out using a spindle-driven ZWICK 1484 at a strain rate of 0.001 s$^{-1}$ and with a pre-force of -50 N. Moreover, both cylinder surface areas were coated with boron nitride to reduce friction. Besides the initial state, we aimed for a total true strain of 12%, 24% and 35%.

## 2.3. Characterisation

Several characterisation methods were used to get an idea of the sheets' microstructures and crystal orientations. For a first macroscopic visual overview, stereographic microscopy (SteREO Discovery.V12, Carl Zeiss AG) was used at low magnification (1-12x) after etching the surface with nitric acid (100ml $HNO_3$, 150ml $H_2O$). Vickers micro-hardness testing (HMV2, Shimadzu Scientific Instruments) with a force of 0.2 N and a holding time of 15 seconds, on the other hand, was used to get an (indirect) overview over the microstructure and deformation state. Mean values were taken from at least 10 measurements. Next, the macroscopic texture was accessed by XRD. Therefore, a commercial three-axis X-ray goniometer (D8 Advanced, Bruker) with a high-resolution area detector (512 × 512-pixel, resolution of 5 °), a 1-mm collimator and an iron anode ($\lambda$ = 1.94 Å) at 30 kV and 25 mA was used. During the experiments, an area of at least 8 x 8 $mm^2$ was scanned, and each frame was measured for 10 seconds. The orientation distribution functions were calculated based on three incomplete pole figures with the help of a self-written MATLAB script using the toolbox MTEX [32]. Furthermore, the A-parameter was calculated based on these results, which is the texture's mean angle deviation of the closest *easy axes* relative to MD [33]. Therefore, the A-parameter is the polycrystalline equivalent to the minimum *easy axis* deviation to MD of the single crystals. Finally, to get more detailed, site-specific information about both the microstructure and microtexture, EBSD measurements were carried out. For this, a dual-beam SEM (Helios Nanolab 600i, FEI Co.) at 20 kV and 11 nA equipped with a Hikari camera (EDAX Inc.) was used. As we are looking at huge grains and aim for a conclusion for the whole sample, an area of > 0.6 $mm^2$ (> 23.4 $mm^2$ for the GO samples) was scanned with a step size of 1 μm (5 μm for the GO samples). Interesting areas were scanned again at a higher magnification and with a smaller step size. OIM Analysis 7 (EDAX Inc.) was used for post-processing and to create different maps. Each scan was cleaned up by grain CI (confidence index) standardization and a single grain dilation iteration (5° tolerance, 2 pixel minimum). Afterwards, only points with a CI value above 0.1 were considered. Mean GND (geometrical necessary dislocation) values were calculated based on the large maps and the following parameters: slip systems – {101}/{112}/{123}<111>; nearest neighbour – $1^{st}$ ($2^{nd}$ for step sizes < 1µm), maximum misorientation – 14° (to exclude high angle grain boundaries).

## 2.4. Electromagnetic Properties

As the size of the grown crystal is limited, a standardized SST cannot be used to measure the global magnetic properties. Therefore, a miniaturised SST was developed together with Dr. Brockhaus Messtechnik GmbH & Co. KG. The minimum sample size for this SST is 25 x 10 mm$^2$ (length x width; length is allowed to be larger; width is fixed) with a yoke leg distance of 22.3 mm, which is in the range of the grown crystals, and two copper winding systems, each having 60 windings. The primary winding is used to generate the external magnetic field and the closer secondary winding to measure the induced field. An MPG 200 unit run with the MPF-Expert software (both from Dr. Brockhaus Messtechnik GmbH & Co. KG) is used to control the SST. To minimize errors from the air flow a self-written numerical compensation is included in the calculation. For a more detailed explanation of the miniaturised SST as well as an additional correcting function applied to make the data comparable to standardized measurements we would like to refer to [34] Generally, an SST is used to determine the hysteresis curve of a material. From this curve magnetic values like coercivity ($H_C$), remanence ($J_r$), maximum susceptibility ($\mu_r$), total iron loss ($P_S$) and needed field strength ($H_{max}$) to reach a certain polarisation ($J_{max}$) can be determined, as shown in Figure 1, where one exemplary quadrant of a symmetric hysteresis curve is shown. Tests were carried out at 10 and 50 Hz. In section 4, normalized values of these magnetic material properties are often used to illustrate differences due to orientation, deformation, or grain morphology.

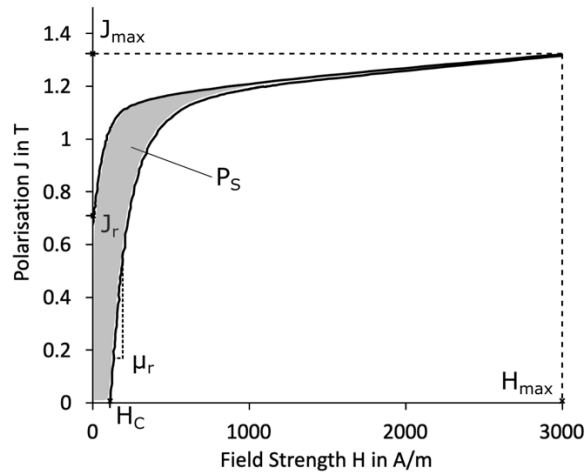

*Figure 1: Exemplary quadrant of a symmetric hysteresis curve, indicating the location where different magnetic material properties are found. Note: $H_C$ – coercivity, $J_r$ – remanence, $\mu_r$ – maximum susceptibility, $P_S$ – total iron loss.*

# 3. Results

## 3.1. Texture & Orientation

The following four figures (Figure 2 to Figure 5) provide information about the textures and crystal orientations of the different sheets. The results of XRD and EBSD agree very well, although the area of the EBSD measurement is much smaller.

The basic idea behind the GO sheets cut in different directions was to obtain orientation reference samples in industrial quality. Since we had 6 samples per direction, the GO samples were also used to check the reproducibility of the SST measurements, which was very good. Industrial GO material has a sharp Goss texture relative to RD, so that by cutting the sheet in certain angles (0°, 45°, 90°) relative to RD we obtained certain axes parallel to MD. Although there is some scatter visible in Figure 2 (a) and (b), GO 0° mostly has a texture with <100> axes parallel to MD, GO 45° with <111> axes parallel to MD and GO 90° with <101> axes parallel to MD (note that these are ordered 0°, 90°, 45° in Figure 2 to achieve consistency with the dominant crystal orientations throughout). The texture intensity is highest for GO 45°, medium for GO 0° and lowest for GO 90°, though an intensity of above 12 can still be considered high. As we are interested in the magnetisation behaviour along MD, we are most interested in the first line of inverse pole figure (IPF) triangles. Thereby, it is noticeable that GO 90° shows a bit more scatter around the desired <101> || MD texture. Furthermore, Figure 2 (c) visualises exemplary unit cell orientations of each group, while also indicating the mean deviation angle of the closest *easy axis* relative to MD, which is low (3.6°) for GO 0°, medium (41.9°) for GO 90° and high (51.9°) for GO 45°. Here it should also be mentioned that GO 90° shows a cube surface rotation (several <100> axes perpendicular to MD) and GO 45° a cube volume rotation. All GO samples have a grain size of above 4 $cm^2$, so there are always a low double-digit number of small angle grain boundaries present per sheet (compare Table 2). The extremely high texture indices underline that we are dealing with oligo-crystals having only minor orientation variations.

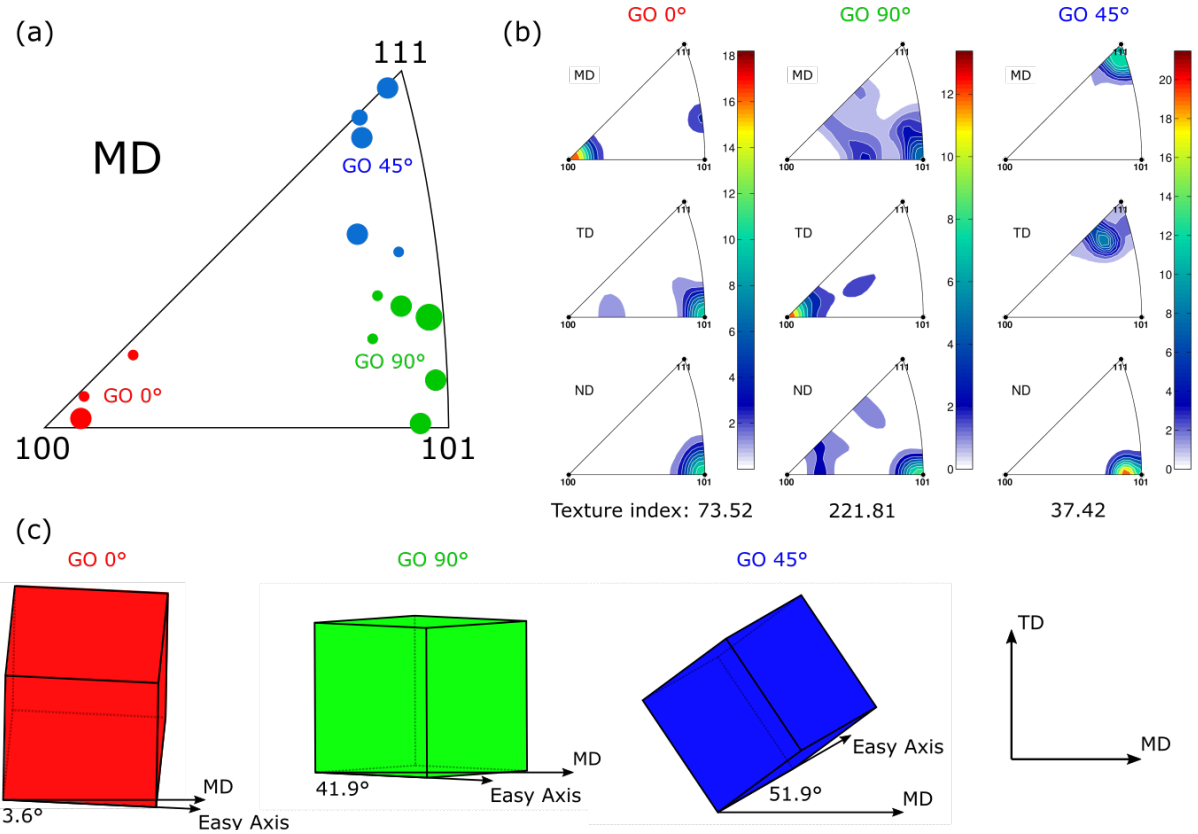

*Figure 2: (a) Microtextures based on exemplary 8.7 x 2.7 mm² EBSD measurements for each GO group, where the size of the circles corresponds to the size of the individual grains; (b) macrotextures and texture indices based on exemplary 10 x 20 mm² XRD measurements for each GO group; (c) schematic, exemplary crystal lattices for each GO group, where the angle is the average deviation of the nearest easy axis relative to MD. Note: colours correspond to a standard MD IPF triangle: red – <100> || MD, green – <101> || MD, blue – <111> || MD.*

Figure 3 shows the results of the grown single crystals. At above 12, the intensity of the XRD measurements is high, and all targeted orientations were achieved with a scatter below 6° in MD. The numbers behind SX indicate which axis is parallel to MD. The texture indices are again extremely high, underlining that we are dealing with single crystals, however, at least two small angle grain boundaries are present in SX100 (compare Table 2, which lowers the texture index to ~87. In addition, the crystal lattices in (c) refer to the points in (a) and show that SX100 has the lowest angle deviation of the next *easy axis* relative to MD (5.7°), SX101 a medium one (40.1°) and SX111 the highest (51.6°).

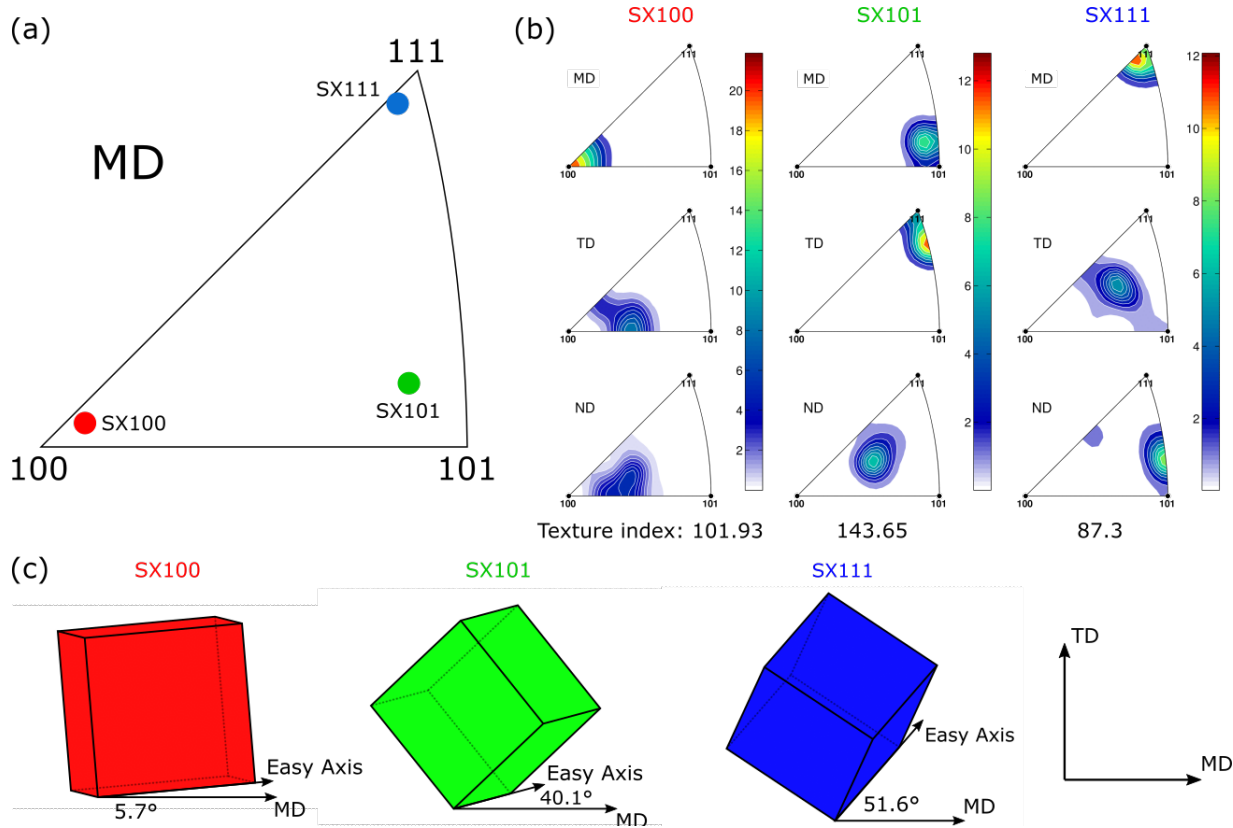

*Figure 3: (a) Microtextures based on exemplary 0.89 x 0.68 mm² EBSD measurements for each single crystal (SX); (b) macrotextures and texture indices based on exemplary 10 x 10 mm² XRD measurements for each SX; (c) schematic, exemplary crystal lattices for each SX group, where the angle is the average deviation of the nearest easy axis relative to MD. Note: colours correspond to a standard MD IPF triangle.*

Figure 4 shows the texture results for the grown bi-crystal in the undeformed as well as the deformed states. The state is indicated by the number behind BX and corresponds to just over 10, 20 and 30% compressive deformation, respectively, see Table 2. The nominal bi-crystal is mainly composed of two grain orientations, labelled BXa/BXa', with <100> near MD, and BXb, which is initially located in the middle of the standard triangle. Grains BXa and BXb are present in all samples and BXa' only occurs in the samples taken to higher compressive strains, BX 2 and BX 3, so that we have tri-crystals with an additional grain in the latter two cases. For a more detailed analysis of the grain morphologies, we refer to Figure 7.

Across the four states, it becomes apparent in Figure 4 (a) that grain BXa initially deviates slightly from <100> || MD along the (100)-(111) axis before it is located exactly in the (100)

corner of the triangle in subsequent samples/states. This resembles a cube texture visualized in (c), with almost no deviation (0.7°) of the *easy* <100> *axis* relative to MD. BXb first has a centred orientation slightly shifted towards <111> || MD before it is located in the (111) corner for BX 1 and BX 2 (see also Figure 7). The mean deviation of the next *easy axis* to MD shown in Figure 4 (c) is 39.3° with a cube volume rotation. For BX 3 no intensity of this grain is visible in the IPF triangle indicating that it was not within the measured XRD area. Grain BXa' has a Goss orientation with a low (3.5°) deviation of the next *easy axis* towards MD. The texture indices are a bit lower than for the single crystals (Figure 3) as at least two grains with very different orientations are present here. What is also striking is that the A-parameter (comparable to *easy axis* deviation of SX) heavily decreases from 36.3° for BX 0 to 18.8° for BX 3, in other words, the mean angle deviation between MD and the next *easy axes* is heavily reduced for BX 3.

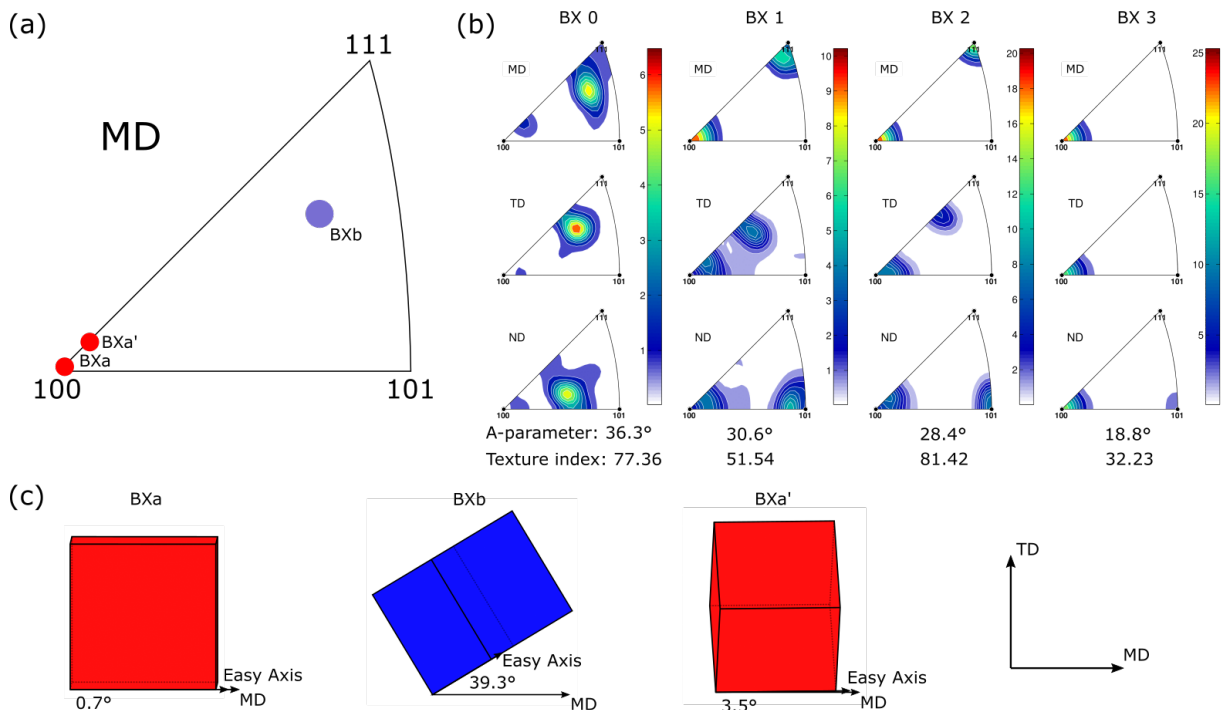

*Figure 4: (a) Microtextures based on 4.850 x 0.68 mm² EBSD measurements for each bi-crystalline (BX) sheet; (b) macrotextures, A-parameter, and texture indices based on 10 x 10 mm² XRD measurements for each BX; (c) schematic, exemplary crystal lattices for each BX grain, where the angle is the average deviation of the nearest easy*

Figure 5 shows the textures of the deformed and undeformed polycrystalline sheets. The texture intensity is generally much lower, also after it nearly doubles for Poly 2 and Poly 3. Nevertheless, the most common directions along compression direction or MD are *easy* <100> *axes*, however, the next strongest axes are the *hard* <111> *axes* especially for Poly 2 and Poly 3. The low texture indices underline that we measured several grains, and the values are several orders of magnitude lower than the values above. Initially (Poly 0), the microstructure consists of equiaxed grains, which become more elongated in TD with successive compression. The mean grain size calculated based on fewer than 100 grains is 177 $\mu m^2 \pm 57.2$ $\mu m^2$ for Poly 0 and 237.6 $\mu m^2 \pm 117.7$ $\mu m^2$ for Poly 3. Therefore, 800.000 to 1.130.000 grains and a multiple of grain boundaries are within the area of the magnetic measurement. No preferred grain boundaries were found. The A-parameter remains relatively constant with increasing deformation and is relatively low with a value that lies between SX100 and SX101 (compare Figure 3).

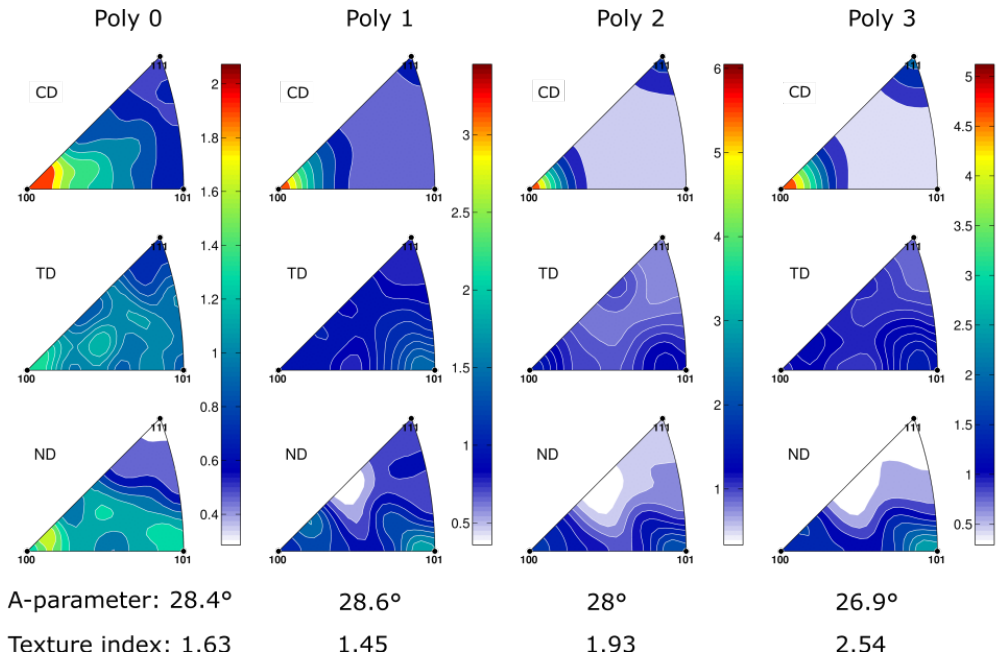

*Figure 5: Macrotexture and A-parameter based on 8 x 8 mm$^2$ XRD measurements for each polycrystal (Poly). Note: CD – compression direction, number behind Poly indicates degree of deformation.*

## 3.2. Deformation structures

Although the stress-strain behaviour is not the focus of this work and the compression testing machine could not be carefully calibrated due to sample scarcity and a non-standard sample geometry (half cylinder), we would like to point out here that SX100 and SX111 had the highest yield strength, followed by BX and SX101 with the lowest. However, SX101 had the highest work hardening rate, followed by BX and SX100 having a similar one and SX111 showed a plateau after 8% true strain. During deformation, cracking sounds were audible for many of the samples.

Figure 6 shows examples of how the different grown single crystals behaved during deformation. SX100 (a) and SX111 (c) exhibit very localised deformation along a line crossing the centre of the cylinder at an angle of ~45°, the latter even shows a second of these lines in the other direction. Both show cracks in the vicinity of these deformation zones. Beyond that, SX100 shows uniform deformation in the EBSD map and a compression axis rotation towards <111> along the (100)-(111) IPF triangle edge. SX101 and SX111 have in common that twins are formed during deformation, some of which can already be seen in the overview images. All twins have a <100> orientation parallel to MD and a 60 ± 1° misorientation towards the matrix. Although it cannot be determined with 100% certainty from a 2D EBSD map, we assume {112}<111> twinning systems. Additionally, SX101 shows a strong compression axis rotation towards <111> along the (101)-(111) IPF triangle edge and SX111 in the other direction, but to a much lesser extent. In the EBSD inlet in Figure 6 (c), we also see how twins taper at a small angle grain boundary, indicating a crucial role of these boundaries during twinning.

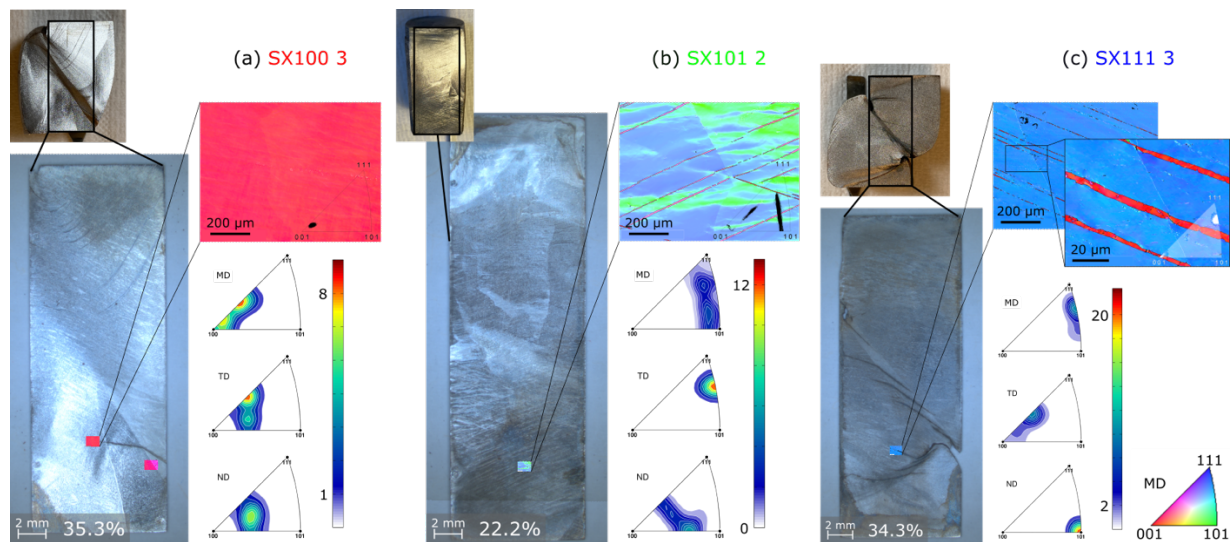

*Figure 6: Overview of the deformed single crystals (SX), each with a photo of the grown cylinder directly after deformation, an optical microscopy picture of the etched sheet cut from the cylinder (dimensions for the miniaturised SST) as well as the macrotexture measured by XRD and the microtexture measured by EBSD. (a) SX100 3 – 35.3% true strain, (b) SX101 2 – 22.2% true strain and (c) SX111 3 – 34.3% true strain.*

As can be clearly seen in Figure 7, the grain boundary in the bi-crystal is not vertical, the grain shares change significantly and in (c) and (d) Grain BXa' appears, as discussed above. From the top and bottom the grain boundary is roughly on the same level, however, along the height of the cylinder it has a curvy trajectory. Nevertheless, the microstructure and texture are well described and can be used to discuss the magnetic properties. Especially in (c), a lot of twinning is seen to take place in grain BXa. Supplementary to the description above, the share of BXb shrinks from 84% in BX 0 to 10% in BX 3. Grain BXa spreads out in the (100) corner upon deformation, while its share increases from 16% in BX 0 to 76% in BX 3. The new grain BXa' more or less adds up to the orientation of grain BXa (at least relative to MD) and has a share of 4% in BX 2 and 13% in BX 3. All grain boundaries are random, mixed high angle grain boundaries (compare Table 2) and are roughly aligned along MD. From this figure it becomes clear why the A-parameter in Figure 4 successively reduces from BX 0 to BX 3, namely because the grain share of <100> grains strongly increases. Furthermore, the reduced texture index of BX 3 also makes sense as we are measuring three instead of two grains during our XRD measurement.

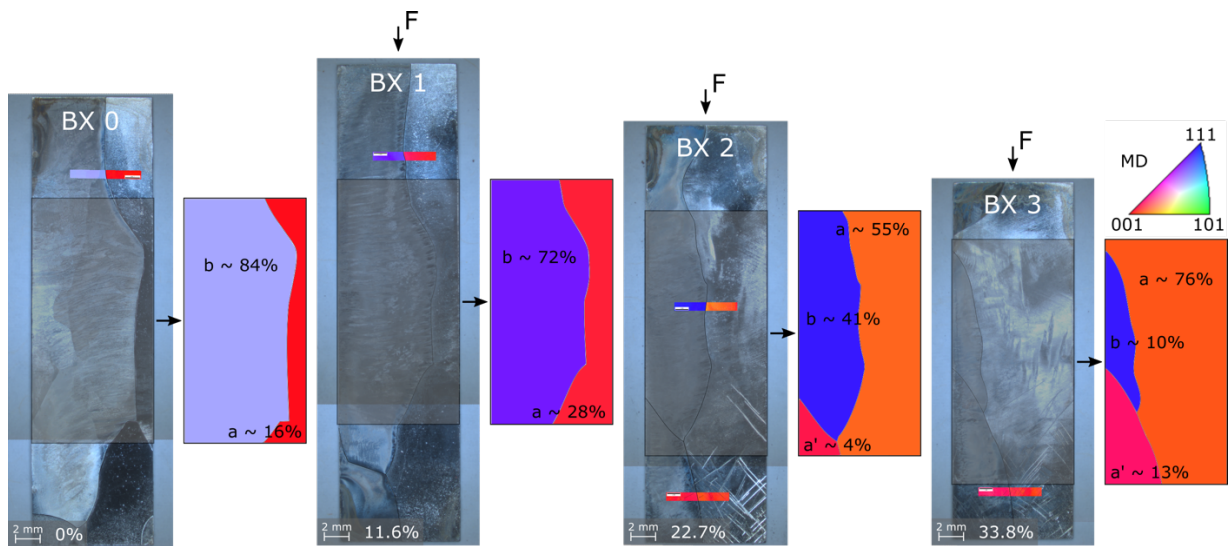

*Figure 7: Overview of the different deformation states of the bi-crystal (BX), each showing the etched cut sheet, the EBSD map(s), the area (grey transparent) subsequently measured in the SST and interpolated IPF colour coded schemes of the SST area based on the EBSD maps and the stereographic optical microscope images to locate grain boundaries with indicated grain shares. Note: F – force.*

Figure 8 adds to the results of the previous figure. In (a) the figure shows how the grain boundary misorientation angle between grain a and b changes with deformation and position through the cylindric bi-crystal. While there is a very distinct 41° orientation change for BX 0 with more or less flat orientations (a little bit of scatter and some small angle grain boundaries) on either side, this angle as well as the orientation distribution within the grains change for BX 1 and BX 2. The grain boundary misorientation angle for BX 1 is still roughly the same, however, in grain a there is a slow misorientation built up to the final grain orientation over 100 μm, which can also be seen in the crystal orientation (CO) map in (b). Grain b, on the other hand, does not show such a built up in (a) as well as in (b). The scatter in BX 2 as well as the grain boundary misorientation angle is much higher and there is a misorientation built up over 200 μm in both grains next to the grain boundary, both of which decrease the grain boundary misorientation angle. Finally, the twin boundary map again highlights that {112}<111> is the active twin system in grain a and a' and that twins do not cross grain boundaries, but often initiate new twins on the other side of the grain boundary.

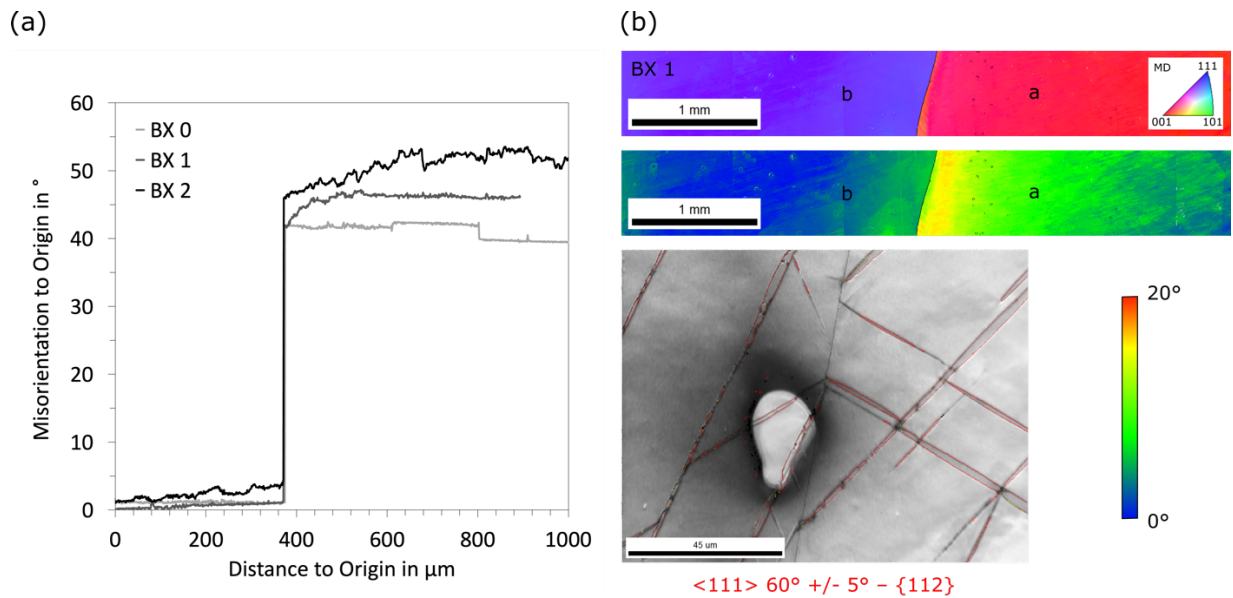

*Figure 8: (a) Misorientation to origin graph based on line scans across the grain boundary between grain a and b of different deformation states of the bi-crystal (BX) and (b) EBSD results of BX strained to 11% showing an IPF map, a crystal orientation (CO) map and a twin boundary map.*

Figure 9 illustrates how the hardness and GND density change with deformation for the grown single and bi-crystals as well as the polycrystal. Initially, the hardness of SX111, SX100 and Poly is a little bit higher (180 HV) than that of SX101 and BX (170 HV), whereby the difference is well within the standard deviation. Upon deformation to 12% the curves of the single crystals split up with SX111 having the steepest slope, SX100 a medium one and SX101 the lowest. Between 12 and 35% the slopes are lower and more or less similar again. SX111 has even higher values than BX and Poly that fall between SX100 and SX111. It should be noted that there are hardness gradients along the growth direction of the crystals (up to 50 HV from top to bottom) as well as higher values (plus 80 HV) in localised deformation zones. Nevertheless, the differences between the sheets remain, also when we include the standard deviations.

The actual dislocation density (composed of SSDs and GNDs) cannot easily be accessed as statistically stored dislocations (SSDs) cannot be measured with EBSD as they do not result in obvious lattice rotation. However, a simplified calculation based on the samples' hardness [35] indicates that the total dislocation density is roughly three orders of magnitude higher than that of the measured GND density. The true dislocation density probably lies in between. Furthermore, the GND densities might be influenced by the resolution of the EBSD pattern (binning) and the used step size for their calculation. Nevertheless, the values can be used for a qualitative comparison, as all were measured and calculated in a similar way.

We first see low GND density slopes up to 12% in Figure 9, before they increase between 12 and 25% and become lower again in the last section, except for the Poly curve which stays steep. Furthermore, Poly has a much higher initial GND density with $6*10^{12}$ m$^{-2}$ compared to $\sim3*10^{12}$ m$^{-2}$ for the grown crystals. BX shows another exception, as here the first section is the steepest one (Figure 9 (b)). Beyond 12% Poly has the steepest slope followed by SX101, SX111, BX and SX100 has the lowest slope. SX100 even shows a decreasing GND density between 25 and 35%, which is however within the common standard deviation [36]. For SX101, the GND density more than quintuples, while for SX100 it just doubles. Grain a of the bi-crystal first has the highest GND density in the bi-crystal before its GND density slightly decreases towards 25% and grain a' takes over without showing such a decrease. Again, these differences are well within the usual standard deviation.

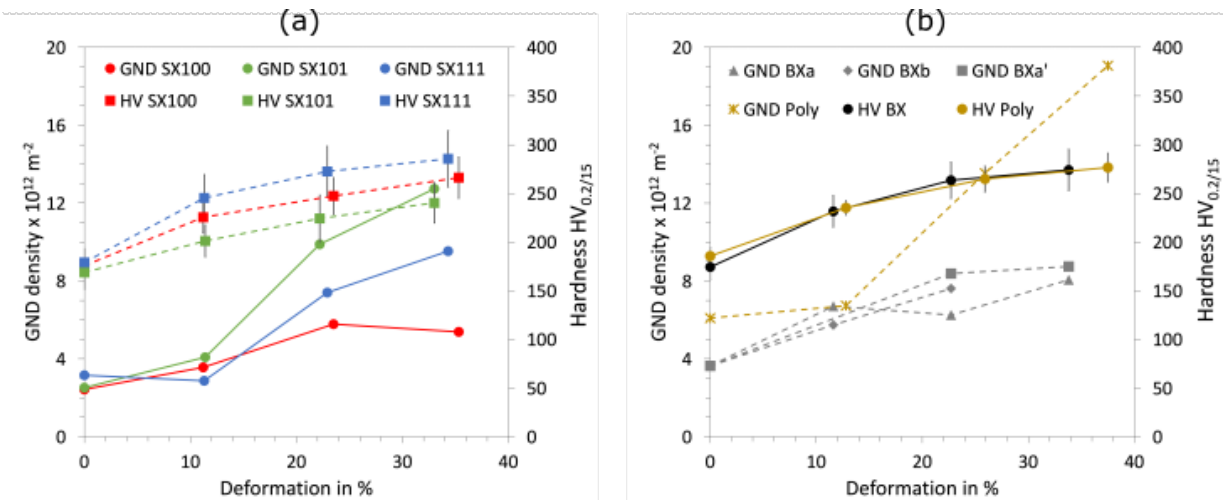

*Figure 9: GND density (dashed lines) and HV (solid lines) versus deformation for (a) the grown, undeformed and deformed single crystals (SX) and (b) the bi-crystals (BX), divided in their individual grains for the GND curves.*

## 3.3. Electromagnetic Properties

In this section, the magnetic properties will be described mainly by means of hysteresis, polarisation, and total iron loss curves. With a view to the hysteresis curves, just one quadrant is shown as the curves are symmetrical. Unless otherwise described, the diagrams refer to tests at 50 Hz.

### 3.3.1. Orientation

Figure 10 gives a first overview of the magnetic properties of the undeformed grown crystals as well as the polycrystalline reference material through polarisation and total iron loss curves. Each data point of these curves is based on its own hysteresis curve, with the distance between the points being 0.1 T. In (a), SX100 and SX111 have similar, high susceptibilities up to ~1 T. At this point SX111 has a distinct knee and it becomes much more difficult to increase its polarisation, while SX100 only shows a slight decrease in slope. SX101, BX and Poly have very similar progressions with knees around 1.25 T, after which the Poly curve lies a little bit above the other two curves. The BX curve is first shifted a little bit to higher field strengths before it approaches the SX101 curve at high polarisations. In (b) all curves show a progressive progression with SX100 having the lowest and Poly the highest total iron loss. SX101, SX111 and BX are very similar and in between SX100 and Poly (closer to Poly above 1 T), whereby SX111 shows discontinuities above 1 T.

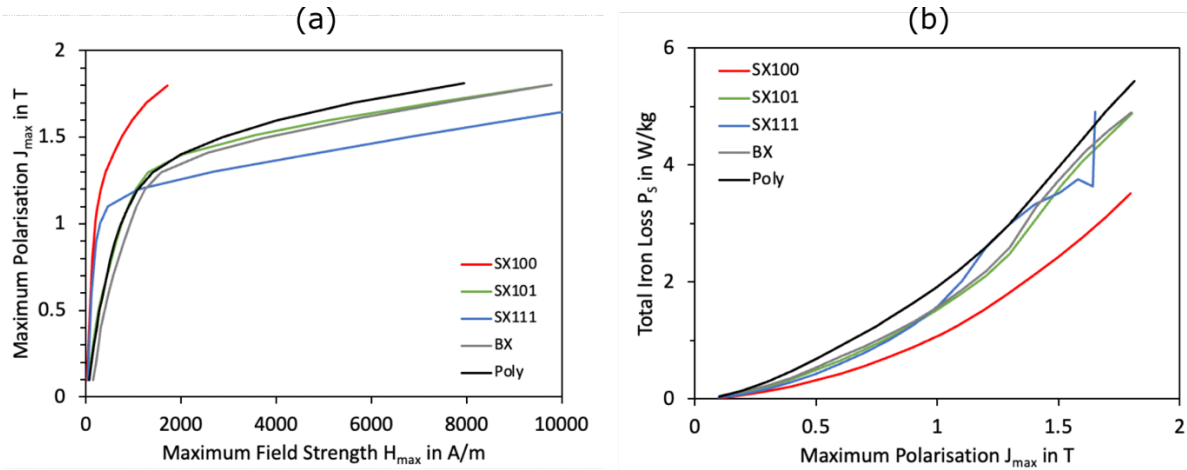

*Figure 10: (a) Polarisation and (b) total iron loss curves of undeformed grown single (SX) and bi-crystals (BX) as well as of the reference polycrystal (Poly).*

### 3.3.2. SX vs GO

In the next figure, differences between GO sheets and the grown single crystals are illustrated. From the hysteresis curves, it can be seen that the susceptibility of the grown material is generally lower than that of the commercial GO material below the knee. Furthermore, the knee is found at higher polarisations for the GO material, whereby GO 45° and SX101 almost lie on top of each other for medium to high polarisations. It is also noticeable that the slope after the knee is similar for GO 90° and SX101 as well as GO 45° and SX111. In addition, SX101 has by far the highest coercivity in the undeformed state.

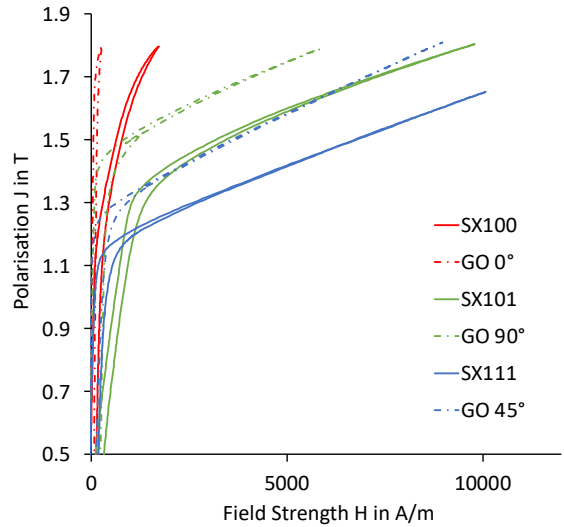

*Figure 11: Excerpt from the symmetric hysteresis curves of SX (solid) and GO (dashed) samples, magnetized to 1.8 T at 50 Hz. The colours correspond to a standard MD IPF triangle.*

### 3.3.3. Deformation

Figure 12 visualises the general trend of how the hysteresis curves change with increasing deformation using SX111 as an example, because the trend is most obvious for this sample group. Coercivity, total iron loss as well as field strength needed to reach a certain polarisation increase with increasing deformation. With the remanence this is not as obvious. Upon initial deformation the remanence increases before it stays more or less the same for SX111 2 and finally decreases again for the third deformation step.

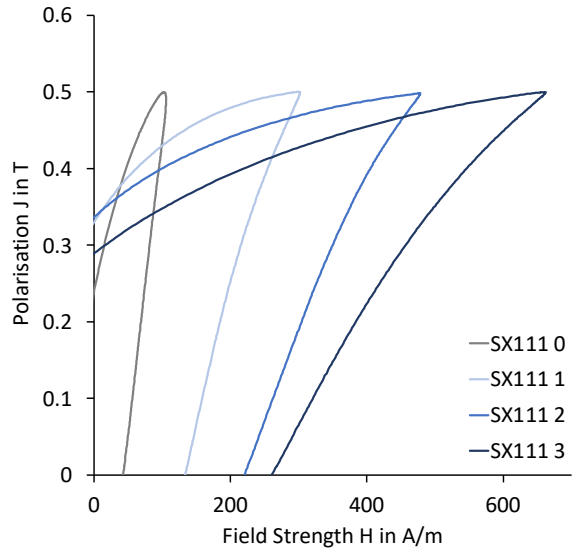

*Figure 12: Excerpt from the symmetric hysteresis curves of the undeformed and deformed SX111 sheets magnetized to 0.5 T at 50 Hz. The last number in the legends indicates the degree of deformation.*

The next figure shows the polarisation and total iron loss curves of the grown single crystals with different degrees of deformation. One exception from the above mentioned more difficult magnetisation with increasing deformation is SX100 1 (Figure 13 (a)), which is easier to magnetise above 1.3 T than the corresponding undeformed sheet. Furthermore, compared to SX100 0, SX100 2 has a knee at a lower polarisation (~0.8 T), SX100 3 at a higher polarisation (~1.4 T) and the curve of SX100 1 does not have a distinct knee. For SX101 it is striking that upon initial deformation (SX101 1) the resulting polarisation curve is a good bit below the initial curve (SX101 0), however, this deterioration does not continue with subsequent deformation steps as SX101 2 and SX101 3 more or less lie on SX101 1. The knees generally become less distinct with deformation. This can also be seen in (e), where the slopes not only decrease in general, but increasingly with deformation. After this increasingly difficult magnetisation at medium polarisations, all SX111 curves approach each other again at high polarisations. There are no exceptions for the total iron losses, the higher the deformation the higher the losses, whereby SX100 has the lowest and SX111 the highest losses. The steps between subsequent curves are more or less equal in size with increasing deformation. All total iron loss curves have a progressive progression with increasing polarisation, while SX111 shows some discontinuities at very high polarisations.

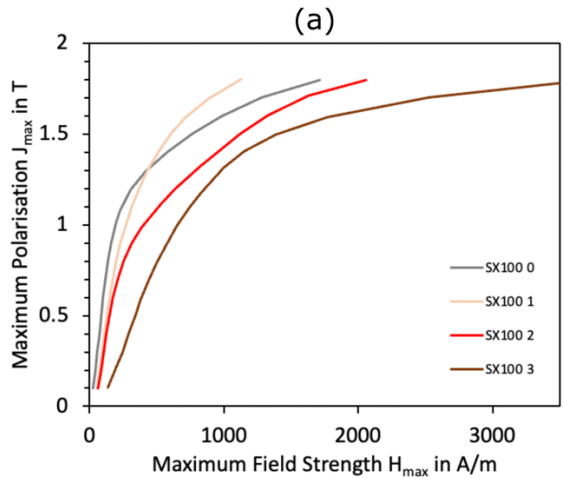

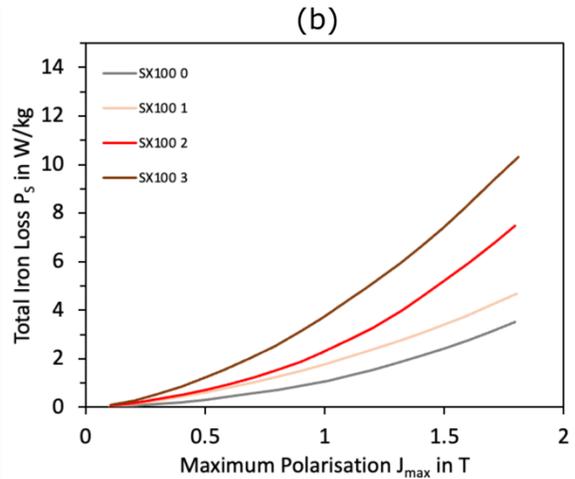

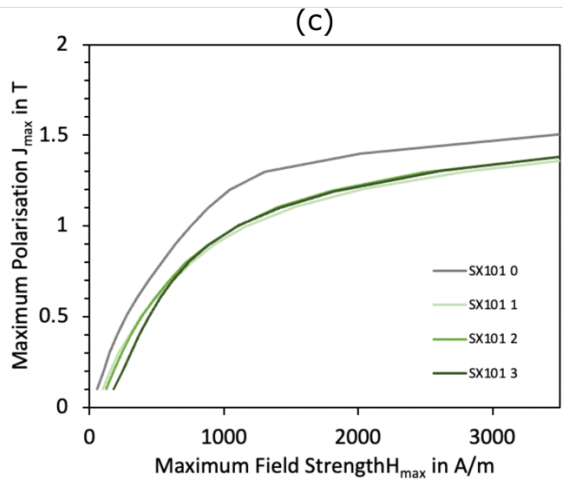

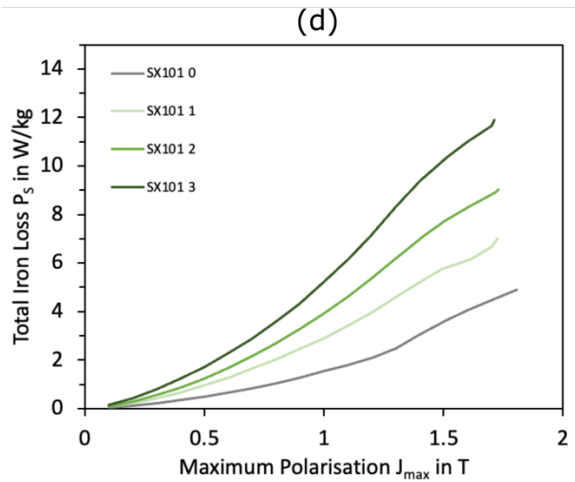

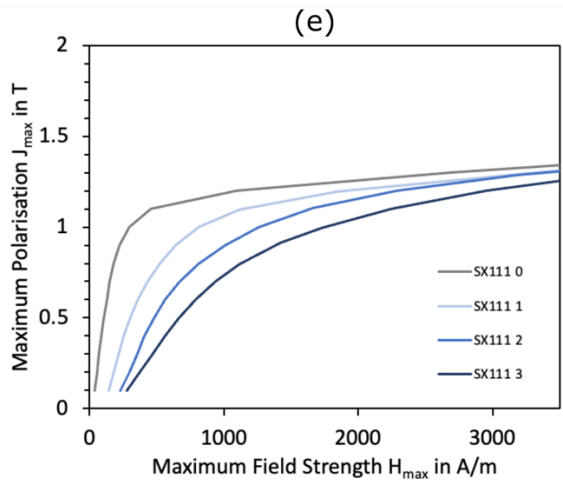

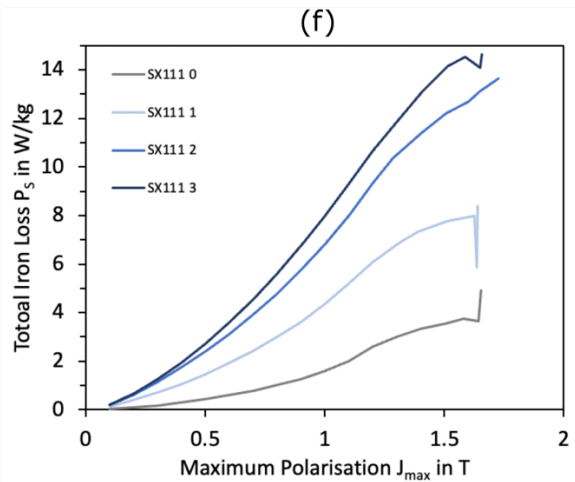

*Figure 13: (a), (c) & (e) Polarisation curves and (b), (d) & (f) iron loss curves of deformed and undeformed grown single crystals (SX). The last number in the legends indicates the degree of deformation and the colour code corresponds to a standard MD IPF triangle.*

### 3.3.4. Frequency

When comparing the magnetic properties tested at different frequencies (10 vs. 50 Hz), the maximum field strength to reach a specific polarisation and the susceptibility show nearly no difference. Iron loss and remanence, on the other hand, have a comparable progression to the coercivity, which is shown in Figure 14. The sequence SX100 > SX101 ≥ BX > SX111 > Poly (nearly constant) is always the same, the curves fan out towards higher polarisations (best seen in (b)) and the distance between the curves as well as the fan magnitude are higher for no or low deformations, whereby the fanning is most obvious for the here shown coercivity. SX111 in (a) must be excluded as it was not possible to get a stable measurement at 10 Hz. In addition, at the high deformation stage in (c) there is a little converging funnel between the first two polarisation steps with Poly increasing from values below one and SX100 decreasing from values above one. It should also be noted that the relative values of iron losses are around 4 to 6 and not around 1 as in this figure.

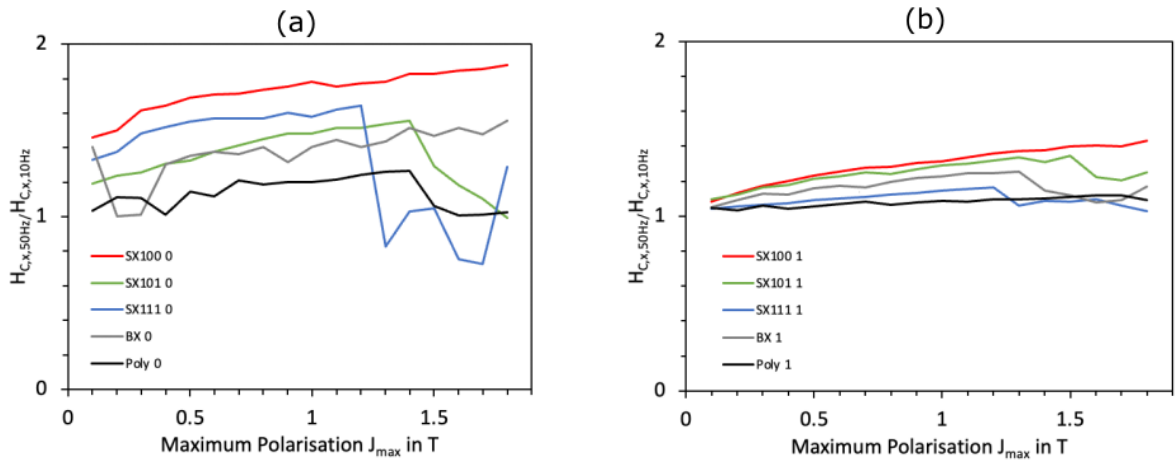

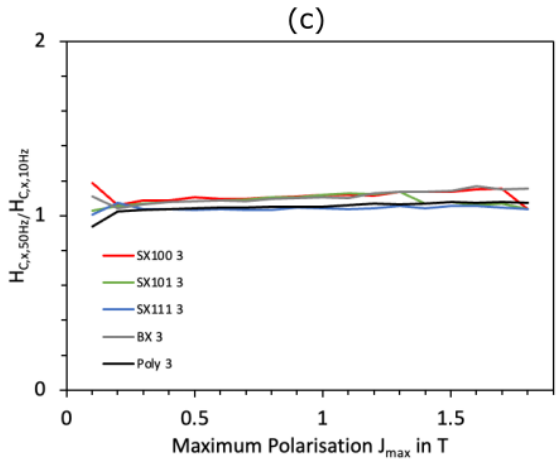

*Figure 14: Comparative coercivity (H_C) tested at 50 and 10 Hz for grown single (SX), bi- (BX) and polycrystalline (Poly) samples. (a) undeformed, (b) deformed to ~11% and (c) to ~33%. The last number in the legends indicates the degree of deformation and the colour code corresponds to a standard MD IPF triangle.*

### 3.3.5. Bi-crystal

Figure 15 shows the first results of the deformed and undeformed bi-crystal. Compared to most other hysteresis curves in this paper, BX 0 has a special S-shape that is otherwise just seen in the 0.5 T GO 90° hysteresis curve. In contrast to Figure 12, the maximum field strength to reach a certain polarisation does not necessarily increase with the degree of deformation here, however, the coercivity and hysteresis area still do. Furthermore, a big jump in remanence can be seen upon initial deformation, however, there is no clear trend for the remanence with further deformation. The overall sequence in the polarisation graph (b) is BX 3 > BX 2 > BX 0 > BX 1, while BX 1 is initially furthest left before crossing all curves at medium polarisations. In (c) a clear correlation between total iron loss and deformation can be seen, whereas BX 3 and BX 2 have the highest losses followed by BX 1 and BX 0. Moreover, the first jump between BX 0 and BX 1 is higher than the subsequent ones.

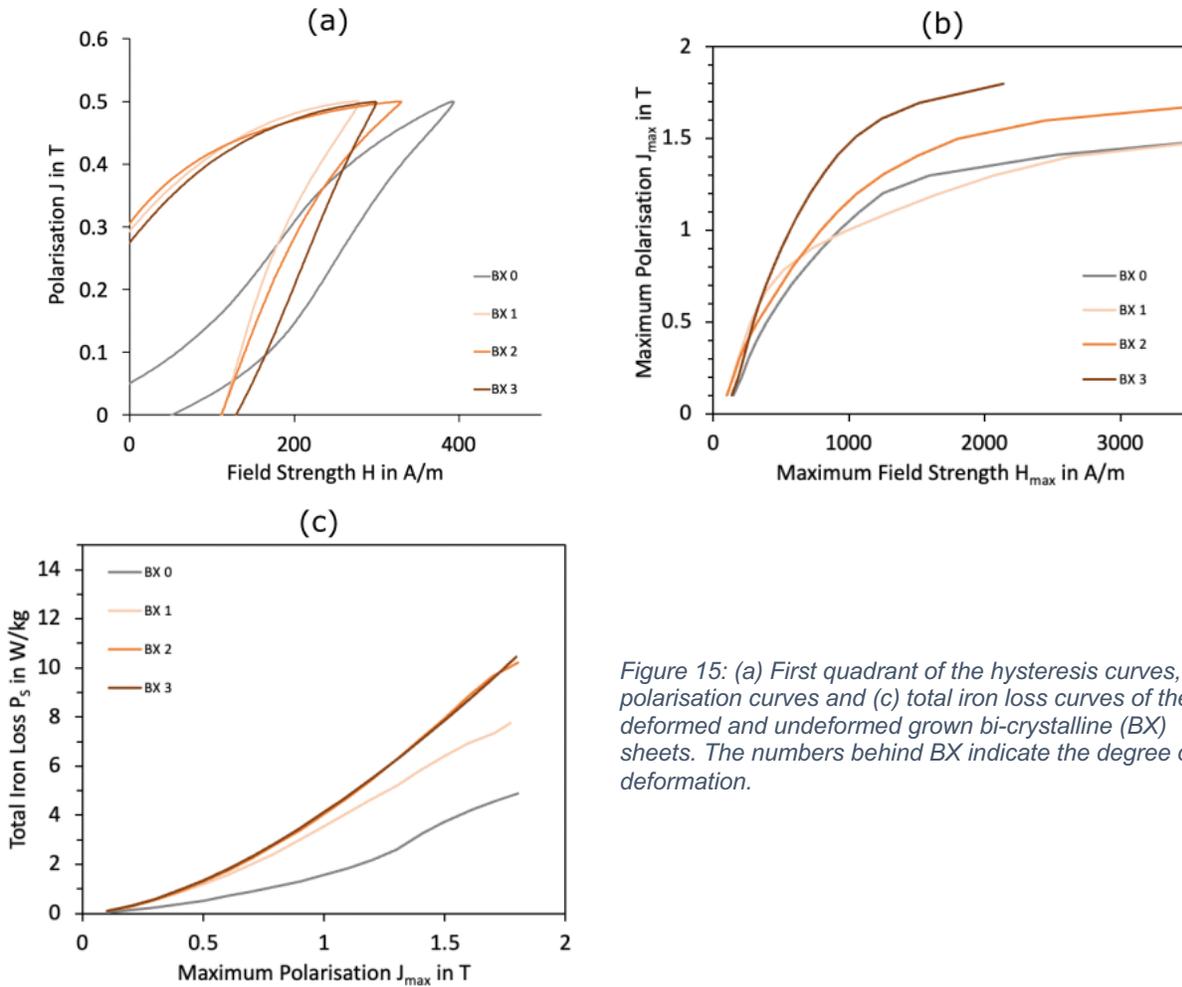

*Figure 15: (a) First quadrant of the hysteresis curves, (b) polarisation curves and (c) total iron loss curves of the deformed and undeformed grown bi-crystalline (BX) sheets. The numbers behind BX indicate the degree of deformation.*

# 4. Discussion

### 4.1. Historic Data

The systematic, fundamental analysis of the orientation dependence of magnetic properties began, among others, with the experiments of Honda and Kaya [2] nearly 100 years ago. Figure 16 compares their results to our investigations on the GO material. We use the GO material here because we have better statistics for these samples and they have slightly better magnetic properties than SX, which we will discuss later. Nevertheless, the general trends are similar for the corresponding GO/SX pairs. The most important finding of Honda and Kaya was that <100> directions are easiest to magnetize, followed by <101> and <111>, with the <111> curve crossing the <101> curve at very high polarisations, which has already been explained in detail in section 1. The main experimental differences to our experiments are that Honda and Kaya used pure iron single crystals, a different setup to measure the magnetic properties (ballistic and torque measurement) and a different sample geometry (rod and ellipsoids).

Three things become clear from Figure 16 (a) – firstly, until now, we were not able to reach the materials' maximum polarisations with our setup, where the curves might converge again; secondly, the division into *easy*, *medium*, and *hard* axis is confirmed by our experiments and thirdly, after the knee, our curves show a steeper slope than those of Honda and Kaya. Regarding the expected crossing of <101> and <111> at very high polarisations, our GO 90° (<101> || MD) curve shows a lower slope than GO 45° (<111> || MD) at 1.8 T, indicating that they will potentially also cross at higher polarisations.

Subfigure (b) reveals that GO 45° (<111> || MD) is easier to magnetize than GO 90° (<101> || MD) at low to medium polarisations before the knees and before the well-known sequence is established. This observation cannot be drawn safely from Honda and Kaya's results as the resolution at low polarisations is too low. However, Pӑltânea et al. [37, 38] have already measured our sequence at low to medium polarisations without giving a detailed explanation of the phenomenon. In the polarisation region in question, favourably oriented domains grow, and entire domains rotate to *easy axes* that are closer aligned to MD. Compared to GO 0° (<100> || MD), the path along *easy axes* from one end of the sample to the other is longest for GO 45° (<111> || MD) and intermediate for 90° (<101> || MD). Furthermore, a higher number of high angle turns with thicker, higher energy domain walls are necessary to reach the other end (GO 0° < GO 90° < GO 45°) [39]. This adds to the explanation why GO 45° and GO 90° are harder to magnetize than GO 0°, however, it does not explain why GO 45° is easier to magnetize than GO 90° at low to

medium polarisations and why GO 45° has a steeper slope after the knee. It appears that domain growth and rotation are more difficult when the relevant *easy axes* are in the two-dimensional MD-ND plane (GO 90°) rather than diagonally in the three-dimensional ND-MD-TD space as in GO 45°. Furthermore, on average, taking all 12 *easy axes* into account, GO 45° has a smaller *easy axis* deviation as it does not have perpendicular *easy axes* towards MD like GO 90° (compare Figure 2 (c)). These perpendicular axes require 90° rotations during hysteresis, which behave differently from 180° rotations [40, 41]. Nevertheless, the higher minimum *easy axis* angle deviation of GO 45° still results in the earlier knee and its name as *hard axis*, although also in the region after the knee the characteristic reversible axes deviation seems to require less energy for GO 45° than for GO 90° as the slope of GO 45° is steeper, the region only begins earlier. The explanation for this is, presumably, also along the lines of the above.

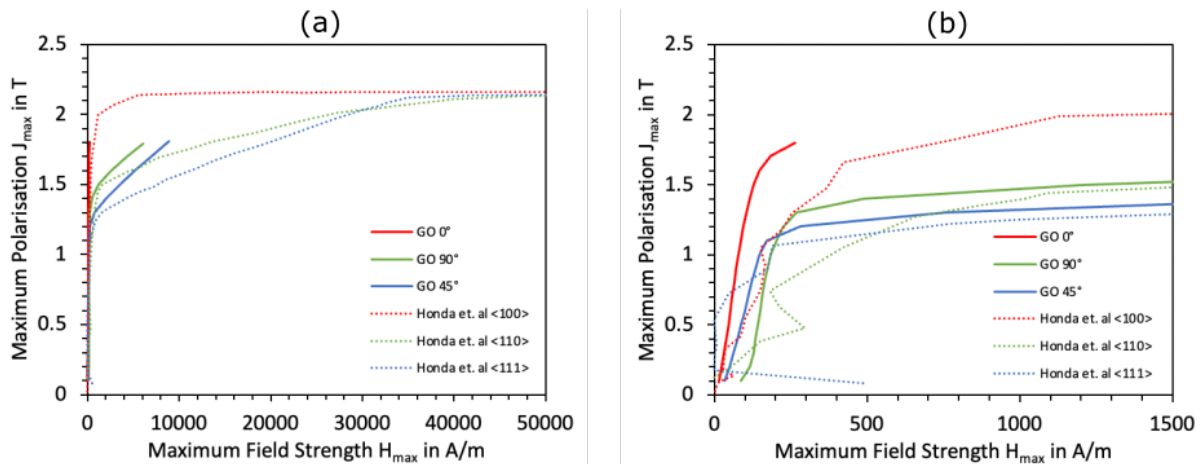

*Figure 16: Polarisation curves showing GO data as well as data from [2], where (b) is a zoom in of (a) towards smaller field strengths.*

### 4.2. Polarisation and Deformation

Before we discuss the correlation of deformation and magnetic properties, we need to briefly consider the deformation behaviour of our samples. To accommodate the strain during compression, at least two deformation mechanisms are active in our material after elastic deformation. On the one hand, there is dislocation movement and multiplication, and, on the other hand, there is deformation twinning, whereby deformation twinning is observed in SX101, SX111 and BXa/a'. The cracking sounds (tin cry [42]) during compression can presumably be correlated to this twin formation, however, the occurrence was not assigned to the individual samples.

Usually, the critical resolved shear stress for deformation twinning is high compared to that of dislocation movement in bcc iron and is therefore not observed in polycrystalline material deformed at room temperature [43, 44]. However, when the mean free path is large and the availability of dislocations and slip systems is limited, deformation twinning has already been observed in bcc iron [44]. With their large grains, all our grown crystals fulfil the first criteria (large mean free path) and after careful crystal growth the dislocation density should be low [45]. Therefore, the availability of slip systems seems to be decisive for the occurrence of deformation twinning, as no twins are observed in SX100, which has many available high Schmid factor slip systems, while twins are observed in SX101 and SX111, which have only a few high Schmid factor slip systems (compare Table 3 and Figure 6). After nucleation, twins expand rapidly until they hit an obstacle (e.g. a grain boundary) that they cannot overcome under the prevailing stress conditions. Meanwhile, they can accommodate relatively high strains in a short amount of time. In Figure 6 (c), three twins are visible that impinge at a low angle grain boundary, which probably was an obstacle during their extension, however, the stress concentrations at the tips of the twins were so high that new twins were directly nucleated on the other side of the grain boundary [46, 47].

The situation is less straight forward for the bi-crystal, where the grains with many high Schmid factor slip systems (BXa, BXa') show twinning and the one with less available slip systems (BXb) does not (compare Table 3 and Figure 7). One explanation could be as follows: upon deformation, the critical resolved shear stress is first reached for dislocation movement in grain a and a' (high Schmid factors). In contrast to the situation in SX100, many dislocations cannot annihilate at the free surface, but pile up at the grain boundaries resulting in lattice rotation (compare Figure 8 (b) – high CO difference in BXa close to the grain boundary), which in turn results in stress concentrations and forest hardening and an exhaustion of mobile dislocations. This results in all conditions for twinning being locally fulfilled.

In the following, both the increasing dislocation density as well as the implication of twins (new orientations, twin boundaries, local distortions) need to be considered while discussing the magnetic properties of the deformed sheets.

SX111 has the highest hardness values (Figure 9), as the activation of dislocation slip is difficult (no high Schmid factor slip systems – compare Table 3). This favours localised deformation, twinning and early cracks (Figure 6 (c)) [48, 49]. More uniform deformation with an even distribution of individual dislocations leads to a lower hardness value (SX101 – Figure 6 (b) & Figure 9). Based on [35, 50] we expect a strong correlation between hardness and dislocation density. The GND density of SX111 is unexpectedly only between those of SX100 and SX101 (Figure 9), however, the localised deformation zones with many dislocations might not have been caught by the EBSD measurement. Furthermore, the GND density slopes presumably remain low towards the first deformation step, as this is where we expect to see the formation of individual SSDs [51], which cannot be accessed by EBSD. Between deformation steps 2 and 3, dislocations pile up and form tangles [43], resulting in lattice distortion/rotation that greatly increases the GND density [52]. At even higher deformations, dislocation walls with larger mean free paths between them form [30], which in turn results in a reduced GND density slope again.

*Table 3: Number of high Schmid factor (SF) slip systems plus the three slip systems with the highest SF and a comment on whether twins, localised deformation or cracks are found in the respective grains or not.*

| Sample | SF > 0.35 | Highest | | 2nd highest | | 3rd highest | | Comment |
|---|---|---|---|---|---|---|---|---|
| SX100 | 20 | (211)[-111] | 0.49 | (312)[-111] | 0.49 | (321)[-111] | 0.48 | localised deformation |
| SX101 | 8 | (-213)[1-11] | 0.5 | (-112)[1-11] | 0.49 | (-123)[1-11] | 0.49 | twins |
| SX111 | 0 | (12-1)[-111] | 0.34 | (23-1)[-111] | 0.34 | (13-2)[-111] | 0.33 | twins, localised deformation, crack |
| BXa | 20 | (2-11)[11-1] | 0.47 | (211)[-111] | 0.47 | (-211)[111] | 0.47 | twins |
| BXb | 8 | (213)[11-1] | 0.47 | (112)[11-1] | 0.46 | (101)[-111] | 0.45 | - |
| BXa' | 20 | (2-11)[11-1] | 0.49 | (3-12)[11-1] | 0.48 | (3-21)[11-1] | 0.48 | twins |

In most of the following figures, individual magnetic properties of deformed sheets are normalized by their respective undeformed counterparts to facilitate the discussion of deformation, orientation and grain boundary related trends.

Figure 17 reveals that the maximum field strength $H_{max}$ to reach low to medium polarisations of SX111 reacts more sensitively and progressively to increasing deformation, while SX100, SX101 and Poly show an increase upon initial deformation before the subsequent curves overlap (SX100 does not completely converge), especially above 0.6 T. Overall, localised deformation appears to be the strongest factor in the observed deterioration of $H_{max}$, as SX111 and SX100, which both exhibit localised deformation behaviour (Figure 6 (a) & (c)), show the strongest deterioration. SX111 additionally shows twinning (Figure 6 (c)), which is likely to contribute further to its deterioration. SX101, which does not show localised deformation (Figure 6 (b)), shows no continuous deterioration with increasing deformation in Figure 17 (b). Here, the initial increase in dislocation density strongly aggravates the domain wall movement between the undeformed state and that after initial straining. This effect is only so visible because no high angle grain boundaries are present that could overshadow this initial deterioration [30]. However, further deformation accompanied with an increase in individual dislocations does not further complicate the polarisation. Therefore, not individual dislocations but accumulations of dislocations have a distinct effect on domain growth and especially domain rotation, as the region around the knee is particularly affected (Figure 13). The strong deterioration around the knee (especially for SX111) could additionally be due to the weakening and spreading of the texture caused by compression axis rotation during deformation (Figure 6 (c)). Although a beneficial <100> direction parallel to MD is formed during twinning, this effect seems to be completely overshadowed by the negative effects of the additional grain boundaries and lattice distortions. In the range of reversible axes deviation above 1.5 T, almost no influence of deformation is observed for most specimens. Exceptions are SX100 1, which has a relative maximum polarisation value below 1 and SX100 3, which still has a relative value way above 1. Both can be related to the present crystal orientation, which is the decisive factor in the reversible axes deviation region. SX100 1 has a slightly closer alignment of <100> to MD, presumably due to fluctuations during crystal growth and cutting, and SX100 3 shows a compression axis rotation away from the *easy axis* (Figure 6 (a)).

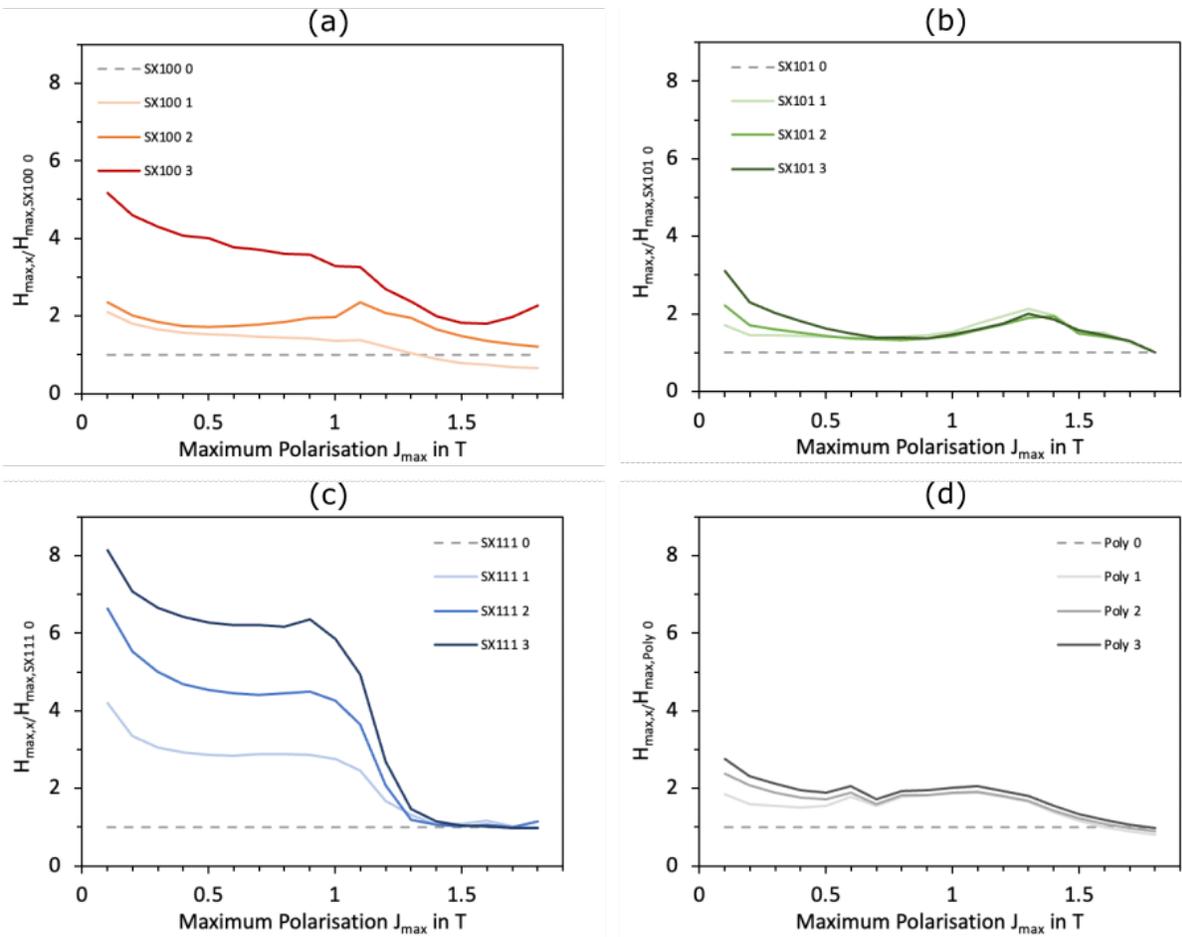

*Figure 17: Relative maximum field strength ($H_{max}$) of the deformed sheets, normalized by the corresponding undeformed samples of (a) SX100, (b) SX101, (c) SX111 and (d) Poly. The colour code corresponds to a standard MD IPF triangle and the last number in the legends corresponds to the degree of deformation.*

### 4.3. Total iron loss $P_S$ & Coercivity $H_C$

The relative total iron loss curves and the relative coercivity curves are very similar in our case. Coercivity gives the opposite magnetic field strength required to demagnetise a material after magnetisation. Therefore, the higher the coercivity, the wider the hysteresis curve and the higher the total iron loss (area enclosed by hysteresis curve). Based on the literature, an increase in coercivity is directly linked to the strength and density of pinning sites in the material [40, 53, 54]. These pinning sites cause normally reversible domain wall movements to become irreversible. Again, SX111 shows the strongest and SX100 the second strongest increase upon deformation (Figure 18). Therefore, not only the movement itself is impeded by the factors mentioned above (localised deformation & twins) but its reversibility naturally is also disturbed. The increase in coercivity, especially for SX111 and SX100, is strongest at low polarisations, suggesting that the barrier to start domain processes (break away from pinning sites) is higher than to maintain them and that the kind of defects play a more important role at these low driving forces, while at higher polarisations the overall defect density is more important (although deformation generally plays a lesser role at high polarisations). Surprisingly, the polycrystal shows the lowest relative increase while having the highest GND density. However, in the undeformed state, it starts with the highest coercivity, which is due to the high number of high angle grain boundaries that act as pinning sites and subsequently partially hide the increase in coercivity due to deformation [30].

We therefore conclude that individual dislocations have the lowest pinning force, followed by twin boundaries and finally localised deformation with networks of dislocations. An additional influence by cracks cannot be excluded.

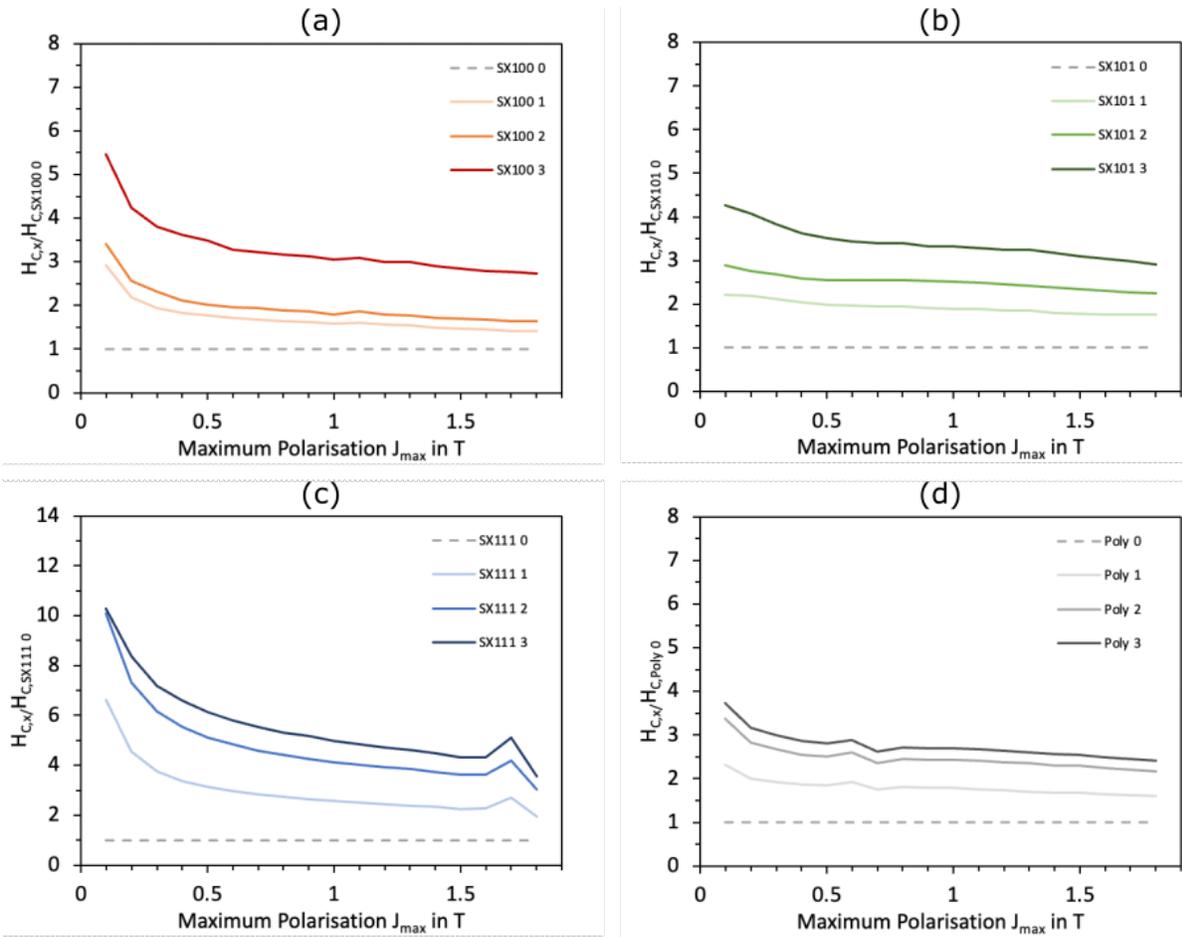

*Figure 18: Relative coercivity ($H_C$) of the deformed sheets, normalized by the respective undeformed samples of (a) SX100, (b) SX101, (c) SX111 and (d) Poly. The colour code corresponds to a standard MD IPF triangle and the last number in the legends corresponds to the degree of deformation.*

Directly comparing the coercivity of SX100 0, SX111 0 and BX 0 at 1.5 T and 50 Hz reveals that BX has the highest value. The biggest difference between the single crystals and BX is the additional grain boundary, which, if other factors (defect density, exact orientation) are left out, increases the coercivity from a mean of 79 A/m ($\frac{SX100+SX111}{2}$) to 110 A/m. Furthermore, this value is only slightly lower than that of the polycrystal (120 A/m), indicating that the initial deterioration (pinning) caused by the introduction of the first high angle grain boundary is more severe than that caused by further grain boundaries.

The remanence is the polarisation that remains after initial magnetisation when the external magnetic field is taken away. In the following two figures, the remanence is normalized by the corresponding maximum polarisation (Figure 19) and by the remanence of the undeformed material, respectively (Figure 20). It is thought that the remanence depends on composition, the thermo-mechanical treatment, and the internal stress distribution [55]. Other factors that are crucial for the magnitude of the remanence are the magnetic anisotropy, the magnetoelastic coupling, and how easy 180° domain walls can nucleate, grow, and move (related to demagnetisation force and domain wall energy/thickness) [55].

Figure 20 shows several interesting as well as unexpected features. First, SX100, SX101, SX111 and Poly all have a value of 0.3 in the beginning before they fan out, which makes sense, since the domain structure does not change much at such low polarisations and similar phenomena take place (domain growth). According to [55] the remanence should be proportional to $H_{max}^2$ in this region. Second, SX111 has a higher remanence than SX100, although according to domain theory, the *easy axis* should give the highest remanence, since the material can reach a stable configuration after magnetisation without much change in the domain structure, which we cannot explain at present. And third, BX has an extremely low remanence, whereby the biggest difference to SX100 and SX111 is the grain boundary. As the magnetic anisotropy is similar and no large internal stress differences are expected (similar hardness value – Table 4), the only reason for the low remanence can be a simplified nucleation and growth mechanism for 180° domain walls in BX around the high angle grain boundary [56]. As no 90 or 180° domain wall can form between the two grains (no perpendicular *easy axes*), we expect "magnetic tension" that promotes domain wall nucleation and demagnetisation. In the polycrystal, on the other hand, the high number of different orientations and grain boundaries prevent a state close to equilibrium and the domain structure "freezes" earlier (domain wall pinning).

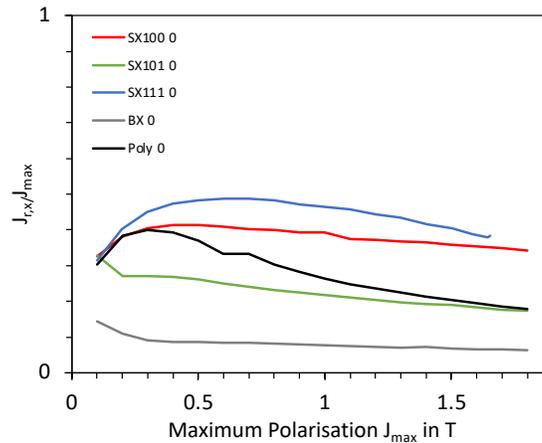

*Figure 19: Relative remanence ($J_r$) of the undeformed grown crystals (SX & BX) as well as the reference polycrystal (Poly), normalized by the respective maximum polarisations ($J_{max}$). The colour code corresponds to a standard MD IPF triangle.*

Usually, under compressive stresses below and above the yield strength of a polycrystalline material with a positive magnetostriction – as <100> in iron – the remanence decreases with increasing stress, while the hysteresis curve is sheared (increased coercivity through domain wall pinning) [57]. Here, we deformed single, bi- and polycrystalline material plastically and measured the magnetic properties ex-situ, that is after deformation, giving mixed results (Figure 20). For the most part, increasing deformation leads to increased remanence, however, for very high polarisations (> 1.2 T), we find a reduced remanence in state 3 of SX100 and SX111 (Figure 20 (a) & (c)). The original idea is that *easy axes* with a positive magnetostriction become harder to magnetize as a result of compressive stresses as the atomic distance is moved away from its optimum [26, 28]. After a previously applied magnetic field is removed, an equilibrium must be established in the magnetic structure, balancing the reduction in magnetostatic energy against the additional energy per domain wall [15]. Here, the assumed formation of individual dislocations in deformation steps 1 and 2 seem to impede the domain wall nucleation and growth process. Furthermore, these dislocations could act as pinning sites for domain walls, resulting in more irreversible domain processes and a "frozen" domain structure closer to the maximum polarisation state [58]. Localised deformation that can be associated with more dense dislocation structures (SX100 3 & SX111 3), on the other hand, seems to facilitate domain wall nucleation and growth as the remanence is reduced again. It is possible that these dense structures serve as nucleation sites. In contrast, further homogeneous deformation between state 2 and 3 (SX101 and Poly) leads to a further increase in remanence (Figure 20 (b) & (d)), random grain boundaries appear to prevent the remanence from decreasing at higher polarisations (Figure 20 (d)) and twins do not seem to have a big influence here, as the two single crystals with twins (SX101 & SX111) do not show similar progressions.

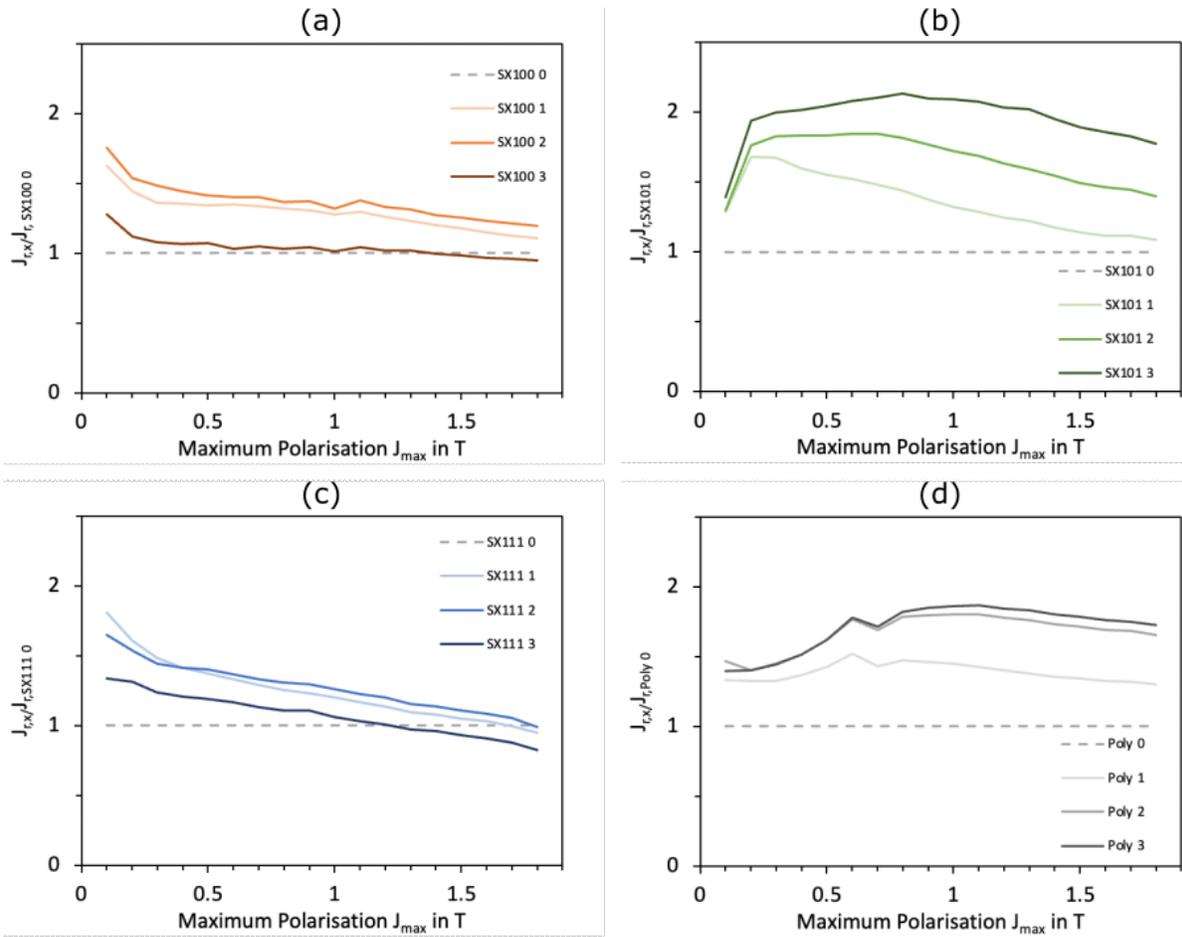

*Figure 20: Relative remanence ($J_r$) of the deformed sheets, normalized by the respective undeformed samples of (a) SX100, (b) SX101, (c) SX111 and (d) Poly. The colour code corresponds to a standard MD IPF triangle and the last number in the legends corresponds to the degree of deformation.*

### 4.5. Susceptibility $\mu_r$

The magnetic susceptibility is the slope in the J/H graph. In other words, this value shows how much the magnetic field is amplified and, for ferromagnetic materials, it changes along the hysteresis curve. In Figure 21, SX100 0, which has the highest susceptibility, was used to normalise the susceptibility of all undeformed samples. SX101, Poly and BX are more or less constant at values between 20 and 40 percent of SX100 and SX111 first has a very high value at low polarisations (80% of SX100), before it drops sharply to having the lowest value at high polarisations (10% of SX100).

When a magnetic field is applied, preferentially oriented domains grow first. In iron the preferred or *easy* direction is <100>, therefore, SX100 has the highest susceptibility. The high value of SX111 at low polarisations is somewhat unexpected as this single crystal has the *hard axes* parallel to MD, however, we already saw in Figure 16 (b) that this crystal has better magnetic properties than SX101 at low polarisations. It seems that preferentially aligned domains in SX111 can grow relatively easily at low polarisations while domain rotations at high polarisations need extremely high energies compared to SX100. Domain growth could be facilitated by the three-dimensional availability of *easy axes* in SX111 as well as the high number of *easy axes* pointing more or less in the right direction (no perpendicular *easy axes* relative to MD). For SX101, Poly and BX, domain growth as well as rotation follow the trend of SX100, but require much more energy. Again, the path along *easy axes* to traverse the sample from one end to the other is longer compared to SX100 and more high angle turns are necessary.

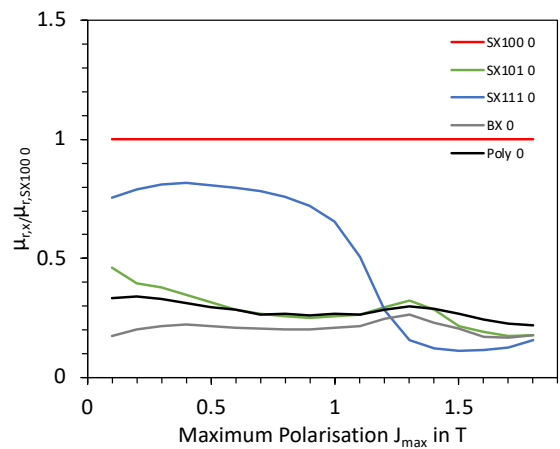

*Figure 21: Relative susceptibility ($\mu_r$) of the undeformed grown crystals (SX & BX) as well as the reference polycrystal (Poly), normalized by the SX 100 0 value. The colour code corresponds to a standard MD IPF triangle.*

Deformation has a big influence on the susceptibility, as dynamic domain wall processes are heavily influenced by magnetoelastic coupling and pinning through dislocations [41]. At low polarisations, all susceptibilities are successively reduced with increasing deformation (Figure 22). The strongest reduction can be found for SX111, which also has the lowest Schmid factors and highest increase in hardness (Table 3 & Figure 9). Combined with the present localised deformation and twinning (Figure 6), we can assume that SX111 has the highest increase in residual stress and stored dislocation density, resulting in this strong deterioration. At higher polarisations, all susceptibility values approach their undeformed value (Figure 22), indicating that the susceptibility in the reversible axes deviation region is not affected by magnetoelastic coupling and dislocations. SX100 1, Poly 1 and Poly 2 even show higher susceptibilities at very high polarisations, which is probably related to variations in crystal orientation. For example, the *easy* <100> || MD texture intensity of Poly 2 more than doubles compared to Poly 0 (Figure 5), due to the activity of certain slip systems and related compression axes rotations. The relatively strong decrease of SX100 3 can be explained with an unfavourable compression axis rotation away from the *easy axis* (Figure 6). After plastic deformation of the polycrystal, we expect mostly compressive residual stresses based on the work of Iordache et al. [40]. With the single crystals, on the other hand, the residual stresses should be lower, as there are no incompatibilities developing between the grains that need to be compensated for, however, some residual stresses could still emerge around twins or any other form of localised deformation and at the top and bottom because of friction during deformation.

Since the decrease in susceptibility is similar between the polycrystal and the single crystals and highest for the first deformation step, we conclude that the reduction is mainly related to domain wall pinning by individual dislocations.

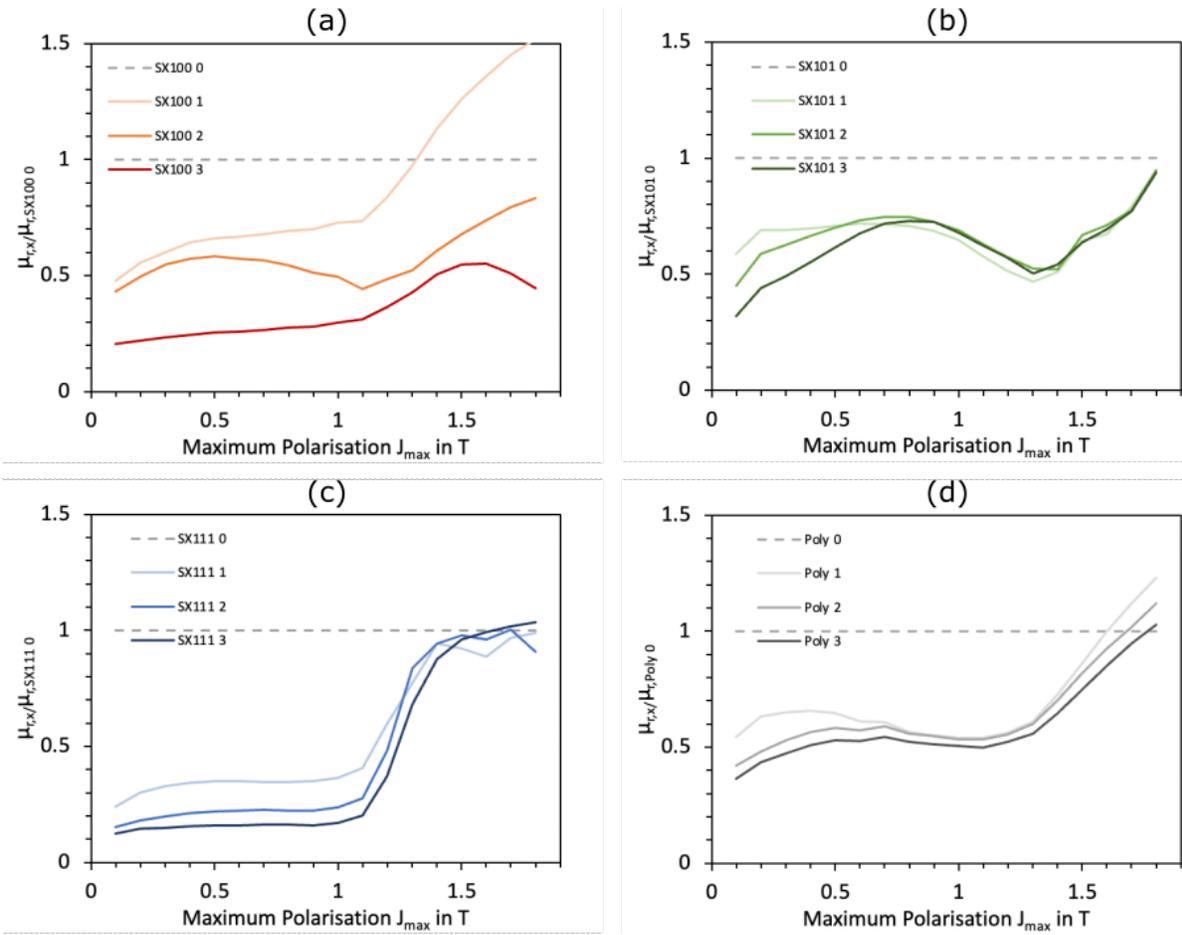

*Figure 22: Relative susceptibility ($\mu_r$) of the deformed sheets, normalized by the respective undeformed samples of (a) SX100, (b) SX101, (c) SX111 and (d) Poly. The colour code corresponds to a standard MD IPF triangle and the last number in the legends corresponds to the degree of deformation.*

### 4.6. Frequency $f$

According to the Bertotti model [59], the hysteresis loss increases linearly with frequency $f$, the eddy current loss squared and the excess loss to the power of 1.5, that is total iron loss $P_{Bertotti} = C_{hyst}\hat{B}^2 f + C_{cl}\hat{B}^2 f^2 + C_{exc}(\hat{B}f)^{1.5}$. Going from 10 to 50 Hz, our total iron loss increases by 4 to 6 times, indicating that the hysteresis loss is dominant in our materials and in the frequency range we tested in. It seems that some magnetisation processes of undeformed samples with few grain boundaries are time dependent and result in a deterioration of specific magnetic properties ($H_C$, $P_S$ and $J_r$) when the frequency is increased (Figure 14) [60, 61]. Thereby, medium to high polarisation processes (domain rotation & reversible axes deviation) are more prone to this deterioration, and the effect is quickly dominated by the effect of deformation or grain boundaries (Poly). SX100, where mostly 180° rotations are necessary during alternating magnetisation, shows the highest frequency dependence, indicating that these rotations need more time/energy.

## 4.7. SX vs GO

The differences between SX and GO (compare Figure 11) can be attributed to the different chemical composition (Table 1) and initial defect density. GO has a slightly higher Si content, which increases the electric resistance and thus decreases the eddy current loss, increases the hardness, and slightly decreases the maximum polarisation [22]. Contrary to this, SX has a higher Mn content which in combination with a higher defect density – the GND density is 2-3 times higher than that of GO – seems to increase the electric resistivity more strongly, as its maximum polarisation is lower than that of GO (CG ~ 1.2 $T_{SX111,knee} * \sqrt{3} = 2.08$; GO ~ 1.3 $T_{GO\ 45°,knee} * \sqrt{3} = 2,25$; [13]). The lower knee positions of SX can be explained through the different maximum polarisation as well as through a slightly different axes alignment (Figure 2 & Figure 3) and the higher defect density. This higher defect density also explains the reduced distinctiveness of the knees, the partially increased iron loss (SX100 – Table 4) and the reduced susceptibility in the low polarisation domain growth and rotation region. Above the knee, in the reversible axes deviation region, the slopes are similar and not affected by the defect density. The remanence might be increased for the GO samples as some domain walls are pinned by small angle grain boundaries, "freezing" the domain structure at a higher magnetostatic energy [62]. On the same basis, small angle grain boundaries do not seem to be a beneficial location for domain wall nucleation.

The discontinuities that occur in the samples with *hard axes* along MD (Figure 10 (b) & Figure 13 (f)) are related to limitations of our setup. For these orientations, we currently reach the limit of the field strength that can be applied, the form factor error rises above 1 % and the hysteresis curve becomes unstable.

## 4.8. BX & Poly

BX and Poly have very similar polarisation curves to SX101 (see Figure 10 (a)). At first glance, this makes sense for BX, as it contains a <100> and a ~<111> grain (Figure 7), resulting in "medium" properties. However, it is surprising that the knee and especially the susceptibility (Figure 21), which is actually lowest for BX out of all grown crystals, are not closer to SX111, as the ~<111> grain is much bigger than the <100> grain. This shows that a thin <100> grain extending over the whole length can already have a big effect on the knee position, while a single high angle grain boundary (with different axes || MD in the adjacent grains) can deteriorate the susceptibility strongly at low to medium polarisations (domain growth/rotation region), possibly through local stray fields (incompatibility of neighbouring *easy axes* resulting in stable domain walls up to high polarisations and a reduced number of available *easy axes*) and to a lesser extent domain wall pinning [62]. This can also be seen for the polycrystal, which has a relatively low A-parameter (28.4°, Figure 5), but a knee and susceptibility similar to SX101, which has an *easy axis* deviation of 40.1° and no high angle grain boundaries (Figure 3). It is striking that the influence of a single or many grain boundaries is so similar.

We therefore conclude that domain growth and rotation are strongly aggravated by high angle grain boundaries and less by potentially present *medium axes* or a high number of domains. In view of the total iron loss, we clearly see the negative (pinning) effect of many grain boundaries, as the polycrystal has the highest loss (see Figure 10 (b)) [63].

An S-shaped hysteresis curve could be explained by an extremely different magnetic behaviour of different microstructural constituents, in the case of BX 0 (Figure 15 (a)), for example, by the orientations of BXa and BXb [64]. Both grains extend over the whole height but not the whole width (Figure 7). Therefore, the right and left sides of the sheet have very different magnetic properties and would individually have very different hysteresis curves. During magnetisation, a mix of both could result in this S-shape.

During deformation, dislocations start to pile-up in front of the grain boundary in BX. This becomes obvious as the grain boundary misorientation angle changes (Figure 8 (a)), which can only be accomplished by GNDs. Furthermore, from Figure 8 (a) and (b) it becomes obvious that BXa has higher dislocation activity, as the change in misorientation angle and lattice rotation are higher, which makes sense as it has a larger number of high Schmid factor slip systems available (Table 3). The grain boundary seems to be an effective obstacle for dislocations, as is expected given its type (Table 2) [65, 66].

Regarding the magnetic properties of BX, orientation, deformation, and the present grain boundary play a crucial role. During magnetisation, orientation is the predominant factor because as the proportion of *easy* <100> grains and deformation increases (Figure 7), a lower field strength is required to reach a certain polarisation (Figure 15 (b)). The effect of deformation becomes obvious when we look at the total iron loss (Figure 15 (c)). Here, the value increases with increasing deformation, whereby initial deformation has the biggest influence and, for BX 2 and 3, part of the increasing loss is compensated by the increasingly beneficial texture (Figure 4). For the hysteresis curves, this means that both the area and the coercivity increase (Figure 15 (a)). Dislocations, twins, and residual stresses effectively disturb the domain wall movements and affect the properties through magnetoelastic coupling. Furthermore, the polarisation of BX 1 (Figure 15 (b)) is mainly disturbed around the knee, underlining the negative effect of newly formed, individual dislocations (and dislocation pile-ups around grain boundaries – Figure 8 (b)) on domain rotation, as other factors (orientation, grain share, grain boundary) remain mostly the same compared to BX 0 (Figure 7). The grain boundary itself seems to heavily reduce the remanence in the undeformed state (Figure 19), as this is the only obvious difference to SX100 and SX111. In this context, it should also be mentioned that the remanence of BX increases heavily to 0.55 - 0.75 for a maximum polarisation of 1.8 T upon deformation. Somehow, the high angle grain boundary with few dislocations in the respective grains seems to facilitate domain nucleation and growth, which prevents a high remanence.

For an easy comparison, all magnetic as well as some mechanical properties are listed in Table 4. M270-35A [67] is an industrial electrical steel grade with a similar thickness and a little bit more Si compared to SX, BX and Poly. GO has a little bit more Si. Si clearly increases the hardness of the material, which can be correlated to an increased brittleness [22]. Beyond that only the undeformed samples with *easy <100> axes* along MD have a lower iron loss at 50 Hz than the industrial grade, however, $H_{max}$ is also undershot by BX 2, BX 3, SX100 and GO 90°. This shows that there is still room for improvement for industrial electrical steel grades.

Table 4: Overview of the samples' hardness values and geometrical necessary dislocation (GND) densities as well as field strengths ($H_{max}$) needed to reach a specific polarisation at 50 Hz and the respective total iron losses ($P_S$). The colour code corresponds to a standard MD IPF triangle.

| Sample | $HV_{0.2/15}$ | $GND * 10^{12}$ | $H_{max}$ at 50Hz & 0.5 T | $P_S$ at 50 Hz & 1 T | $H_{max}$ at 50Hz & 1.5 T | $P_S$ at 50 Hz & 1.5 T |
|---|---|---|---|---|---|---|
| GO 0° | 209.2 ± 1.1 | 1.07 | 46.255 | 0.713 | 128.066 | 1.448 |
| GO 90° | 209.2 ± 1.1 | 1.007 | 142.061 | 1.607 | 1204.297 | 3.332 |
| GO 45° | 207.3 ± 1.5 | 1.014 | 87.297 | 1.372 | 3558.628 | 3.97 |
| SX100 0 | 175.6 ± 6.9 | 2.425 | 85.38 | 1.101 | 761.4 | 2.43 |
| SX100 1 | 225.5 ± 14.1 | 3.567 | 129.589 | 1.79 | 604.105 | 3.417 |
| SX100 2 | 247.3 ± 18.3 | 5.785 | 147.01 | 2.24 | 1122.848 | 5.211 |
| SX100 3 | 266.1 ± 22.9 | 5.387 | 341.905 | 3.695 | 1385.303 | 7.404 |
| SX101 0 | 168.8 ± 17.3 | 2.524 | 271.7 | 1.535 | 3547.713 | 3.635 |
| SX101 1 | 201.1 ± 16.3 | 4.082 | 384.164 | 2.894 | 5449.624 | 5.733 |
| SX101 2 | 224.2 ± 24 | 9.889 | 388.158 | 3.898 | 5296.221 | 7.715 |
| SX101 3 | 240.3 ± 20.5 | 12.732 | 441.266 | 5.225 | 5585.1343 | 10.353 |
| SX111 0 | 179 ± 14.1 | 3.161 | 105.7 | 1.588 | 6829.82 | 3.532 |
| SX111 1 | 245.3 ± 23.9 | 2.875 | 302.078 | 4.36 | 7440.404 | 7.784 |
| SX111 2 | 272.6 ± 25.8 | 7.403 | 478.424 | 6.815 | 6978.764 | 12.202 |
| SX111 3 | 285.4 ± 29 | 9.536 | 662.054 | 7.93 | 7150.656 | 14.172 |
| BX 0 | 174.3 ± 12.5 | 3.645 | 394.3 | 1.572 | 3715.954 | 3.72 |
| BX 1 | 231.6 ± 16.2 | 6.22 | 279.121 | 3.522 | 3869.37 | 6.426 |
| BX 2 | 263.5 ± 18.6 | 7.356 | 330.239 | 4.03 | 1795.699 | 7.954 |
| BX 3 | 274.1 ± 21.2 | 8.458 | 299.53 | 4.091 | 1055.88 | 7.947 |
| Poly 0 | 185.7 ± 8.7 | 6.093 | 288.8 | 1.895 | 2845.946 | 3.965 |
| Poly 1 | 235.2 ± 7.3 | 6.736 | 447.469 | 3.961 | 3312.778 | 6.721 |
| Poly 2 | 264.8 ± 13.2 | 13.536 | 495.542 | 5.681 | 3477.99 | 9.814 |
| Poly 3 | 276.6 ± 14.6 | 19.063 | 547.503 | 6.417 | 3823.445 | 11.192 |
| M270-35A [67] | 190 | - | - | 1.1 | 2500 | 2.7 |

# 5. Conclusion

In this work we investigated the fundamental relationships between deformation, crystal orientation and grain boundaries on the one hand and magnetic properties on the other hand in standard electrical iron silicon steels. To analyse and quantify the effects of individual orientations and grain boundaries, macroscopic single and bi-crystals were first grown using the vertical Bridgman-Stockbarger method. In a next step, these crystals were deformed to different strains, showing different deformation behaviour (twinning, local deformation, dislocation propagation). Subsequently, we analysed the microstructure with the help of optical microscopy, hardness measurements, XRD measurements, SEM imaging and EBSD. As these grown crystals are still small (25 x 10 x 0.3 mm$^3$) compared to industrial material, we miniaturised a Single-Sheet-Tester to measure their magnetic properties. When trying to determine which feature or mechanism influences which magnetic property, one must consider the energy balance between magnetostatic, exchange and anisotropy energy, which is influenced by domain nucleation, growth and rotation, as well as reversible axes deviation. All in all, the most important findings from this new approach and from the correlation of the microstructural with the magnetic results are the following:

- The miniaturised SST provides reasonable and reproducible results (e.g. axis-dependent polarisation curves from the literature confirmed; similar GO samples show similar results) and can resolve the influence of crystal orientation, deformation, and grain boundaries of single to oligo-crystals.

- The resolution of magnetic single crystal measurements at low polarisations has been improved compared to [2], showing a deviating sequence of the three standard axes (*easy > hard > medium*) and an additional crossing between the *hard* and *medium axis* before the knees. This has been explained by geometrical considerations of the *easy axes* orientations and mean angle deviations in 2 and 3D, which also explains the susceptibility distribution of the axes.

- Domain wall pinning, which is associated with the degradation of magnetic properties at low to medium polarisations (reduced maximum polarisation at specific field strengths; higher Coercivity and iron loss), is strongest in localised deformation regions with potentially dense dislocation structures, followed by twinned regions with lattice distortion and additional grain boundaries and finally individual dislocations.

- Relatively speaking, introducing the first grain boundary (bi-crystal) has a stronger negative effect on the coercivity than additional ones (polycrystal).

- The susceptibility is mainly deteriorated by individual dislocations.

- On the contrary to the literature, the remanence increases for deformed samples with twins and homogeneous deformation in our experiments and decreases as soon as localised deformation appears. This might indicate that dense dislocation structures and high angle grain boundaries act as domain nucleation sites, which is supported by the extremely low remanence of BX in the undeformed state. For the polycrystal, on the other hand, the pinning strength of the high number of grain boundaries seems to dominate.

- The strongly varying magnetic properties of the grains in BX 0 lead to an S-shaped hysteresis curve and already a small grain extending over the entire height of the sample can have a large influence on the magnetic properties, in this case especially on the knee position.

- The Influence of visible cracks in SX111 on the magnetic properties is surprisingly small

This new approach of targeted crystal growth in combination with miniaturised SST measurements opens up a very wide field. In the future, different grain boundaries can be investigated, especially those where the grains have similar axes parallel to MD, in order to completely exclude the influence of crystal orientation on the magnetic properties. Moreover, deformed and undeformed single to oligo-crystals can be tested at different frequencies and excitation waveforms. Furthermore, this approach can be complemented by analyses of the domain wall structures using magnetic force microscopy, Bitter imaging or TEM investigations.

## Acknowledgements

This work was funded by the Deutsche Forschungsgemeinschaft (DFG, German Research Foundation): 255711070 (IMM), 255713208 (IEM); and carried out in the research unit 1897 "FOR 1897 Low-Loss Electrical Steel for Energy-Efficient Electrical Drives".

# Appendix A. Supplementary material